\documentclass[trackchanges,twocolumn]{aastex701}
\usepackage{amsmath}
\usepackage{booktabs}
\usepackage{multirow}

\begin{document}

\title{Foreground Groups and Host-Galaxy Geometry in the Dispersion-Measure Scatter of Fast Radio Bursts}

\author[orcid=0009-0007-0367-0631, gname=Jiale, sname=Wang]{Jiale Wang}
\affiliation{Department of Astronomy, Xiamen University, Xiamen 361005, China}
\email{wangjiale@stu.xmu.edu.cn}

\author[orcid=0000-0002-2853-3808, gname=Taotao, sname=Fang]{Taotao Fang}
\affiliation{Department of Astronomy, Xiamen University, Xiamen 361005, China}
\email[show]{fangt@xmu.edu.cn}
\correspondingauthor{Taotao Fang}


\begin{abstract}
Localized fast radio bursts (FRBs) show substantial scatter in dispersion measure (DM) about the mean cosmic DM--redshift relation, also known as the Macquart relation. Both foreground structure and host galaxies contribute to this scatter. In this work, we examine foreground tracers and host galaxy properties using a parent catalog of 136 localized FRBs with measured DMs and spectroscopic host redshifts. Among 77 FRBs within the Gfinder footprint, DM excess relative to a fixed Macquart mean increases with foreground group count excess, with a Spearman coefficient of 0.459. Adding the foreground term to a mean host baseline reduces the leave-one-out root mean square error (RMSE) and mean absolute error (MAE) in extragalactic DM by \(13.0\%\) and \(7.5\%\), respectively. For 20 of these FRBs with optical host axis ratio measurements, adding a foreground term calibrated on the other 57 FRBs to the host shape model reduces the observer-frame leave-one-out RMSE from \(76.3\) to \(64.0~\mathrm{pc\,cm^{-3}}\) and MAE from \(54.2\) to \(50.2~\mathrm{pc\,cm^{-3}}\). And the in-sample residual standard deviation in the host rest frame decreases by \(20.2\%\). Idealized galaxy models based on the Milky Way electron density distribution produce DM changes with inclination comparable to the fitted amplitude under the adopted assumptions.
\end{abstract}


\section{Introduction}
\label{sec:introduction}

At low redshift, a substantial fraction of cosmic baryons is thought to reside in diffuse ionized gas that is difficult to detect directly \citep{Shull2012ApJ...759...23S}. Fast radio bursts (FRBs) probe this gas through their dispersion measures (DMs), which trace the free-electron column density along the line of sight \citep{McQuinn2014ApJ...780L..33M}. The observed DM includes contributions from the Milky Way, intervening cosmic gas, and the host galaxy and local environment of the burst \citep{Petroff2019A&ARv..27....4P}. The Macquart relation describes the increase in mean cosmic DM with redshift, allowing localized FRBs to probe the cosmic baryon content \citep{Deng2014ApJ...783L..35D,Zheng_2014,Macquart2020Natur.581..391M}. Localized FRBs show substantial scatter about this mean relation \citep{Baptista2024ApJ...965...57B}, and the relative contributions of foreground structure and host environments to this scatter remain uncertain.

Intervening cosmic structure introduces additional DM variations between sight lines. Studies of individual FRB sight lines have inferred DM contributions from foreground halo gas ranging from small to substantial \citep{Simha2021ApJ...921..134S,Simha2023ApJ...954...71S}. At the population level, excess cosmic DM has been found to correlate with foreground galaxy overdensity, with a stacked excess extending to megaparsec projected separations from foreground galaxies \citep{Hussaini2025ApJ...993L..27H}. Gas associated with galaxy groups, clusters and filaments can also contribute to these variations \citep{Lanman2026ApJ..1003....5L,Mo2025ApJ...995..183M}.

Cosmological simulations suggest that host contributions can dominate the variance in extragalactic DM \citep{Zhu2021ApJ...906...95Z}. For disk galaxies, the host contribution depends on the ionized-gas distribution, the burst position and the viewing inclination, with higher typical DMs expected for more inclined systems \citep{Xu2015RAA...15.1629X}. A study of DMs corrected for the Galactic disk contribution found a correlation with host specific star-formation rate, but no significant association with inclination \citep{Li2025arXiv250701270L}. \citet{Glowacki2025PASA...42..157G} reported an association between optical host axis ratio and the absolute extragalactic Faraday rotation measure. More recently, \citet{Marnoch2026arXiv261013062M} reported a relation between inferred host DM and optical axis ratio using an expanded set of host-shape measurements. Their analysis subtracts the mean cosmic DM contribution without applying additional sightline-specific foreground corrections. Such foreground variations can also affect the scatter about the host-shape--DM relation and its predictive performance.

In this work, we examine foreground tracers and host-galaxy properties to investigate DM scatter about the Macquart relation. Section~\ref{sec:methods} describes the data, observable definitions, and statistical methods. Section~\ref{sec:results} presents the foreground group relation and assesses the additional value of foreground information for host-shape modeling and DM prediction. We discuss the physical implications and limitations in Section~\ref{sec:discussion} and summarize our conclusions in Section~\ref{sec:Conclusions}.

\section{Data and Methods}
\label{sec:methods}
\subsection{FRB Sample and DM Excess}
\label{sec:methods-frb}

We compiled 136 localized FRBs from the public \texttt{FRBs/FRB} database \citep{Prochaska2025zndo..16763432P} and published measurements, requiring spectroscopic redshifts for the adopted host galaxies and a largest adopted localization error scale \(\sigma_{\rm loc}<20''\). To limit potential DM contributions from dense local environments, we excluded five FRBs associated with persistent radio sources (PRSs) \citep{Chatterjee2017Natur.541...58C,Moroianu2026ApJ...996L..16M,Niu2022Natur.606..873N,Bruni2024Natur.632.1014B,Bruni2025A&A...695L..12B}, leaving 131 FRBs for the foreground-tracer and host-property selections. Appendix~\ref{sec:appendix-frb-catalog} provides the catalog (Table~\ref{tab:frb-sample}), exclusion flags, and localization details; Figure~\ref{fig:sky-coverage} shows the sky distribution. We also tested the joint exclusion of three FRBs with uncertain host associations (Appendix~\ref{sec:appendix-robustness}).

For each FRB, we calculated the extragalactic DM as
\begin{equation}
DM_{{\rm exgal},i}
=DM_{{\rm obs},i}
-DM_{{\rm MW,disk},i}
-DM_{{\rm MW,halo},i},
\label{eq:methods-dm-exgal}
\end{equation}
where the Galactic disk and halo DM contributions were estimated with NE2025 and YT2020, respectively \citep{Ocker2026ApJ..1002....3O,Yamasaki2020ApJ...888..105Y}. We assigned these contributions independent, untruncated Gaussian uncertainties centered on the model estimates, with standard deviations equal to 50\% of those estimates.

The Macquart relation describes the mean cosmic DM as a function of source redshift \citep{Macquart2020Natur.581..391M}. We calculated this mean as
\begin{equation}
\begin{aligned}
&\langle DM_{\rm cosmic}(z)\rangle_{\rm Macquart}
=\frac{3cH_0\Omega_{\rm b}}{8\pi Gm_{\rm p}} \\
&\qquad\times\int_0^z
\frac{f_{\rm d}(z')(1-Y_{\rm He}/2)(1+z')}{E(z')}
\,\mathrm{d}z'.
\end{aligned}
\label{eq:methods-macquart-mean}
\end{equation}
Here, \(E(z)=H(z)/H_0\), \(f_{\rm d}(z)\) is the fraction of cosmic baryons in diffuse ionized gas, and \(Y_{\rm He}=0.25\) is the helium mass fraction. We assumed fully ionized hydrogen and helium and adopted the Planck18 cosmology \citep{Planck2020A&A...641A...6P}, with \(H_0=67.66~\mathrm{km\,s^{-1}\,Mpc^{-1}}\) and \(\Omega_{\rm b}=0.04897\), retaining its full expansion history, including radiation and massive neutrinos.

Following \citet{Macquart2020Natur.581..391M}, we calculated the diffuse-baryon fraction as
\begin{equation}
f_{\rm d}(z)=1-
\left(1+\frac{M_{\rm rem}}{M_\star}
+\frac{M_{\rm ISM}}{M_\star}\right)
\frac{\rho_\star(z)}{\rho_{\rm b,0}},
\label{eq:methods-diffuse-fraction}
\end{equation}
where \(M_\star\), \(M_{\rm rem}\), and \(M_{\rm ISM}\) denote the masses in stars, stellar remnants, and neutral interstellar gas, respectively. The quantities \(\rho_\star(z)\) and \(\rho_{\rm b,0}\) are the comoving stellar mass density and the present-day mean baryon density, respectively.

We retained the Macquart et al. mass ratios \(M_{\rm rem}/M_\star=0.30\) and \(M_{\rm ISM}/M_\star=0.38\), extending their constant neutral-ISM-to-stellar mass ratio from \(0\leq z\leq1\) to higher redshift. We evaluated \(\rho_\star(z)\) by cubic interpolation of the adopted stellar mass-density table based on \citet{Madau2014ARA&A..52..415M}.

We held the cosmic mean fixed in the primary analysis and defined the DM excess as
\begin{equation}
DM_{{\rm excess},i}=DM_{{\rm exgal},i}
-\langle DM_{\rm cosmic}(z_i)\rangle_{\rm Macquart},
\label{eq:methods-dm-excess}
\end{equation}
with both \(DM_{\rm exgal}\) and \(DM_{\rm excess}\) in the observer frame. This excess includes contributions from the host galaxy and the local environment of the burst, together with sightline-to-sightline deviations of the cosmic DM from its mean.

We tested sensitivity to Galactic subtraction by replacing NE2025 with NE2001 or YMW16 \citep{Cordes2002astro.ph..7156C,Yao2017ApJ...835...29Y}, and separately adopting a halo contribution of \(30~\mathrm{pc\,cm^{-3}}\). Galactic contributions retained 50\% uncertainties. Guided by the approximately 30\% uncertainty in the non-diffuse baryon fraction discussed by \citet{Macquart2020Natur.581..391M}, we multiplied \(1-f_{\rm d}(z)\) by 0.7 or 1.3 at each redshift and recomputed the Macquart integral. We also repeated the foreground analysis for \(z\leq1\) to assess sensitivity to extending the constant neutral-ISM-to-stellar mass ratio to higher redshift. The foreground model was refitted for each modified input or sample.

\subsection{Foreground and Host Observables}
\label{sec:methods-foreground}

To investigate how foreground structure traces DM variations, we used Gfinder galaxy and group catalogs \citep{Yang2021ApJ...909..143Y,Wang2022ApJ...936..161W}, 2MRS galaxy and group catalogs \citep{Huchra2012ApJS..199...26H,Lambert2020MNRAS.497.2954L}, and MCXC and Planck PSZ2 cluster catalogs and their combined catalog \citep{Piffaretti2011A&A...534A.109P,Planck2016A&A...594A..27P}. We also used galaxy-density tracers constructed from DESI DR1 LSS and WISE$\times$SuperCOSMOS \citep{Ross2025JCAP...01..125R,Bilicki2016ApJS..225....5B}, and the reconstructed 2M++ density field \citep{Carrick2015MNRAS.450..317C}.

For object counts and weighted sums, foreground objects were required to satisfy \(z_{\rm obj}\geq0.005\) and \(z_{\rm obj}+0.05<z_{\rm FRB}\). The lower limit excludes very nearby structures with large angular apertures; the redshift buffer reduces contamination from objects near the FRB redshift and foreground misclassification due to photometric-redshift uncertainties. For DESI DR1 LSS and WISE$\times$SuperCOSMOS, we included only complete redshift shells whose upper edges satisfy this foreground selection.

The catalogs differ in sky coverage and radial extent, so each tracer was evaluated for all FRBs with valid measurements under the adopted angular and radial selections. We examined object-count excesses, weighted foreground sums, and density-field summaries. The Gfinder catalogs are based on Legacy Surveys DR9 imaging \citep{Dey2019AJ....157..168D}; 77 of the 131 retained FRBs lie within their coverage. The main foreground results focus on the Gfinder group-count excess (Section~\ref{sec:results}). The definitions, apertures, background estimates, selection requirements, and comparisons for the other tracers are given in Appendix~\ref{sec:appendix-foreground-exploration}.

For Gfinder groups, we adopted the catalog centers, redshifts, halo masses, and memberships, retaining systems with at least two members. Following the halo-radius definition of \citet{Yang2021ApJ...909..143Y}, we calculated the physical group radius as
\begin{equation}
R_{180\mathrm{m}}=
\left(\frac{3M_h}{4\pi\,180\,\rho_{\rm m,0}}\right)^{1/3}
\frac{1}{1+z_g},
\label{eq:methods-r180m}
\end{equation}
where \(M_h\) and \(z_g\) are the group halo mass and redshift, and \(\rho_{\rm m,0}\) is the present-day total matter density in Planck18, including massive neutrinos. This radius encloses a mean density 180 times the cosmic mean matter density at \(z_g\). We counted groups whose centers lie within a projected distance of \(1.5R_{180\mathrm{m}}\) from the sightline in the primary analysis. We repeated the analysis with apertures of \(R_{180\mathrm{m}}\) and \(2R_{180\mathrm{m}}\) to assess sensitivity to the aperture choice.

To measure foreground counts relative to the typical background for each catalog and foreground-redshift range, we used random sightlines within the angular regions adopted for the Gfinder, 2MRS, and cluster comparisons. Each observed count was compared with 10,000 random counts using the same foreground-redshift selection and aperture definition. We defined the direct count excess as
\begin{equation}
\Delta N_i=N_{{\rm obs},i}
-\operatorname{median}_{j}N_{{\rm rand},ij},
\label{eq:methods-group-count-contrast}
\end{equation}
where \(N_{{\rm obs},i}\) and \(N_{{\rm rand},ij}\) are the counts along the FRB sightline and its \(j\)th random control.

To examine host contributions, we assembled stellar masses, star-formation rates (SFRs), mass-weighted stellar ages, and stellar metallicities from \citet{Gordon2023ApJ...954...80G} and \citet{Sharma2024Natur.635...61S}; effective radii, S\'ersic indices, and projected FRB offsets from \citet{Gordon2025ApJ...993..119G} and \citet{Sharma2024Natur.635...61S}; gas-phase metallicities from \citet{Yamasaki2026ApJ...997L...6Y}; and H\(\alpha\)-based SFRs and gas-phase metallicities from \citet{Muller2026ApJ..1001..118M}. The stellar-population SFRs are averaged over 100 Myr. We also calculated specific SFR, stellar-mass and SFR surface-density proxies, and offsets normalized by the host effective radius. The definitions of these derived quantities and the measurement selections are given in Appendix~\ref{sec:appendix-derived-host-quantities}.

Our host-shape sample is based on the GALFIT measurements presented by \citet{Glowacki2025PASA...42..157G} and expanded most recently by \citet{Marnoch2026arXiv261013062M}. We adopted the axis ratios \(q=b/a\) and statistical fit errors compiled in Table~1 of the latter study. These measurements were obtained from single-S\'ersic fits to the optical light distribution. The 20 hosts within the Gfinder footprint form our axis-ratio sample and are listed in Table~\ref{tab:host-analysis-inputs}. The propagated axis-ratio uncertainties are formal fitting errors and may not capture systematic differences in imaging or structural modeling. We used \(1-q\) to describe projected host shape; larger values correspond to more elongated shapes. For alternative shape measurements, we converted the \(r\)-band AutoProf and Photutils inclinations of \citet{Bhardwaj2024Natur.634.1065B} to approximate axis ratios using their assumed intrinsic axis ratio of 0.2 and analyzed the two measurement sets separately.

\subsection{Statistical Modeling and Prediction}
\label{sec:methods-statistics}

We examined associations of DM variations with the candidate foreground tracers and host properties using Spearman rank coefficients and two-sided \(p\) values, without adjustment for multiple comparisons. Each comparison used the FRBs with the required measurements. Foreground tracers were compared with observer-frame \(DM_{\rm excess}\).

To quantify the relation between foreground group-count excess and \(DM_{\rm excess}\), we fitted the 77 Gfinder-covered FRBs with
\begin{equation}
DM_{{\rm excess},i}\sim\mathcal{N}\!\left(
\frac{H}{1+z_i}+kX_i,\,
s_i^2+\sigma_{\rm F}^2
\right),
\label{eq:methods-main-regression}
\end{equation}
where \(X_i\) is the direct Gfinder group-count excess within \(1.5R_{180\mathrm{m}}\). The parameter \(H\) is the host-rest-frame mean-host normalization, \(k\) is the observer-frame foreground coefficient, and \(\sigma_{\rm F}\) is additional observer-frame scatter. The Galactic-DM variance is \(s_i^2=(0.5DM_{{\rm MW,disk},i})^2+(0.5DM_{{\rm MW,halo},i})^2\). We compared this model with a mean-host baseline having \(k=0\), fitting \(H\) and \(\sigma_{\rm F}\) separately in each model.

After accounting for foreground variations, we examined whether host-galaxy properties could explain part of the remaining DM scatter. We compared these properties with the corrected host-rest-frame residual
\begin{equation}
R_i=(1+z_i)
\left(DM_{{\rm excess},i}-\overline{k}_{\rm cal}X_i\right),
\label{eq:methods-host-residual}
\end{equation}
where \(\overline{k}_{\rm cal}\) is the posterior-mean foreground coefficient. For each property, corrected and uncorrected comparisons used the same FRBs with Gfinder coverage and the required measurements. These exploratory host-property comparisons used a common foreground coefficient calibrated on all 77 FRBs.

We then reassessed the axis-ratio relation using a foreground term recalibrated on the 57 FRBs outside the 20-host axis-ratio sample. For these same 20 hosts, we compared four models to assess the separate and combined contributions of foreground information and host shape, fitting the measured \(DM_{\rm excess}\) directly:
\begin{equation}
DM_{{\rm excess},i}\sim\mathcal{N}\!\left[
kX_i+\frac{A(1-q_i)+B}{1+z_i},\,
s_i^2+\frac{\sigma_{\rm H}^2}{(1+z_i)^2}
\right],
\label{eq:methods-host-regression}
\end{equation}
where \(A\) and \(B\) are the slope and intercept of the host-rest-frame shape relation. The additional host-rest-frame scatter \(\sigma_{\rm H}\) can include cosmic contributions untraced by the foreground catalog. The four models are a mean-host baseline (\(A=k=0\)), foreground plus mean host (\(A=0\)), host shape (\(k=0\)), and their combination. Each model had its own fitted intercept and scatter; models containing a foreground term used the external calibration.

Foreground calibration used independent zero-mean Gaussian priors with standard deviation \(1000~\mathrm{pc\,cm^{-3}}\) for \(H\) and \(k\). Conditional host fits used the same priors for \(A\) and \(B\). The additional scatter parameters \(\sigma_{\rm F}\) and \(\sigma_{\rm H}\) had half-normal priors with scale \(500~\mathrm{pc\,cm^{-3}}\). We integrated over the linear coefficients analytically and evaluated the scatter posteriors by numerical quadrature.

Host fits propagated input uncertainties through 10,000 Monte Carlo realizations. Models containing a shape term drew each axis ratio independently from a Gaussian centered on its adopted value, with its reported standard deviation and truncation to \(0<q\leq1\). Models containing a foreground term drew \(k\) from the fixed 57-FRB calibration posterior, with each draw shared by all hosts. We combined the separately normalized conditional host posteriors with equal weight and report posterior means and equal-tailed posterior intervals.

We quantified in-sample scatter using the sample standard deviation of residuals about each model's full-sample posterior-mean relation, with denominator \(N-1\). Residuals were evaluated in the observer frame for foreground fits and in the host rest frame for host fits. These dispersions include measurement uncertainty.

Leave-one-out predictions were evaluated against observer-frame \(DM_{\rm exgal}\) using the root-mean-square error (RMSE) and mean absolute error (MAE), adding the fixed cosmic mean to each predicted \(DM_{\rm excess}\). Each foreground fold fitted the other 76 FRBs; each host fold fitted the other 19 hosts, retaining the external 57-FRB calibration where applicable. Host predictions used the held-out source's adopted central axis ratio where applicable and were averaged over uncertainty realizations. All models within each prediction comparison used the same sources. The leave-one-out analysis refitted model parameters while keeping the predictor definitions and functional forms fixed.

We used the \texttt{scientific-autoresearch} workflow \citep{Wang2026ScientificAutoresearch} with GPT-5.6-sol and Kimi K3 to compare candidate foreground observables and host galaxy properties under human supervision. GPT-5.6-sol also assisted with language editing. The authors reviewed the final manuscript and take responsibility for its content.

\section{Results}
\label{sec:results}

\subsection{Foreground Galaxy Groups Trace DM Excess}
\label{sec:result-foreground}

\begin{figure}[t]
\centering
\includegraphics[width=\columnwidth]{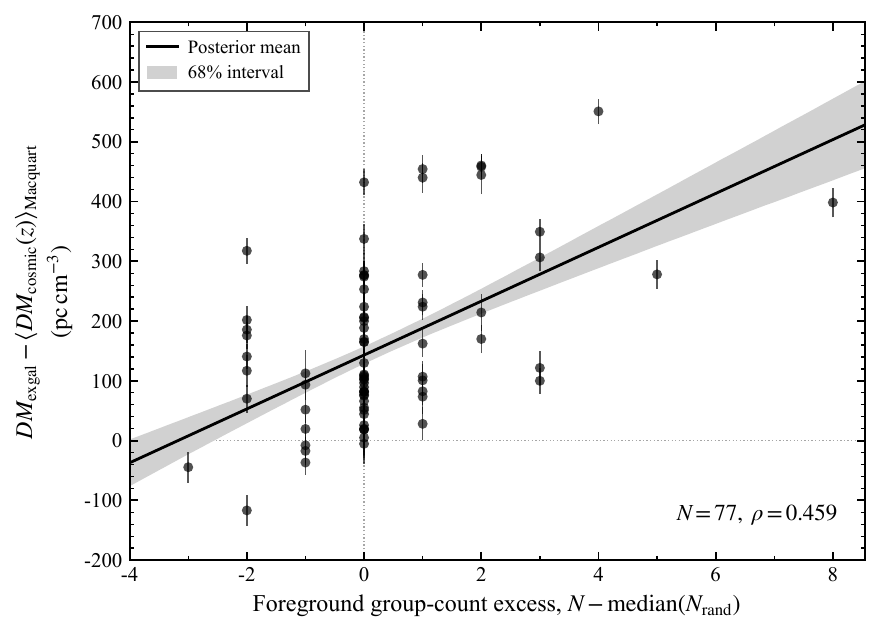}
\caption{Correlation between observer-frame \(DM_{\rm excess}\) and foreground group-count excess for 77 FRBs. Each point represents one FRB. Groups are counted when their centers lie within a projected distance of \(1.5R_{180\mathrm{m}}\) from the sightline, and the median count along random sightlines within the catalog coverage and with the same foreground redshift selection is subtracted to obtain the count excess. The line shows the posterior-mean relation averaged over the sample redshifts, with its pointwise 68\% posterior interval shaded. Error bars show the 16th--84th percentile ranges propagated from Galactic-DM uncertainties.}
\label{fig:primary}
\end{figure}

Among the foreground tracers examined, we find a positive correlation between the Gfinder group-count excess and \(DM_{\rm excess}\) (Figure~\ref{fig:primary}). Across the 77 FRBs within catalog coverage, the Spearman coefficient is \(\rho=0.459\), with a nominal two-sided \(p=2.72\times10^{-5}\). The foreground fit gives \(k=45.1~\mathrm{pc\,cm^{-3}}\), with a 95\% posterior interval of \(27.8\)--\(62.3~\mathrm{pc\,cm^{-3}}\), corresponding to the observer-frame DM increase per unit group-count excess. Comparisons with the other foreground tracers are presented in Appendix~\ref{sec:appendix-foreground-exploration}.

Adding the group-count term to the mean-host baseline reduces the observer-frame in-sample residual standard deviation from \(146.2\) to \(125.2~\mathrm{pc\,cm^{-3}}\). The model also gives smaller leave-one-out prediction errors for \(DM_{\rm exgal}\): the RMSE decreases from \(147.1\) to \(128.0~\mathrm{pc\,cm^{-3}}\), and the MAE from \(115.1\) to \(106.4~\mathrm{pc\,cm^{-3}}\), corresponding to reductions of \(13.0\%\) and \(7.5\%\), respectively.

Replacing the median random background with the mean preserves the positive foreground association, giving \(\rho=0.411\) and \(k=44.2~\mathrm{pc\,cm^{-3}}\), with a 95\% posterior interval of \(26.4\)--\(61.9~\mathrm{pc\,cm^{-3}}\). The leave-one-out RMSE and MAE are \(129.4\) and \(105.8~\mathrm{pc\,cm^{-3}}\), respectively, both below those of the mean-host baseline.

The 95\% posterior interval for \(k\) remains above zero across the other tested model and sample variations, including imaging-matched random sightlines (Table~\ref{tab:robustness-summary}a). Localization uncertainties produce only small changes in the rank correlation: the 2.5th--97.5th percentile range is \(0.447\)--\(0.473\). Omitting individual FRBs gives \(\rho=0.436\)--\(0.498\).

\subsection{Foreground Information Improves DM Predictions Beyond Host Shape}
\label{sec:result-host}

\begin{figure*}[t]
\centering
\includegraphics[width=\textwidth]{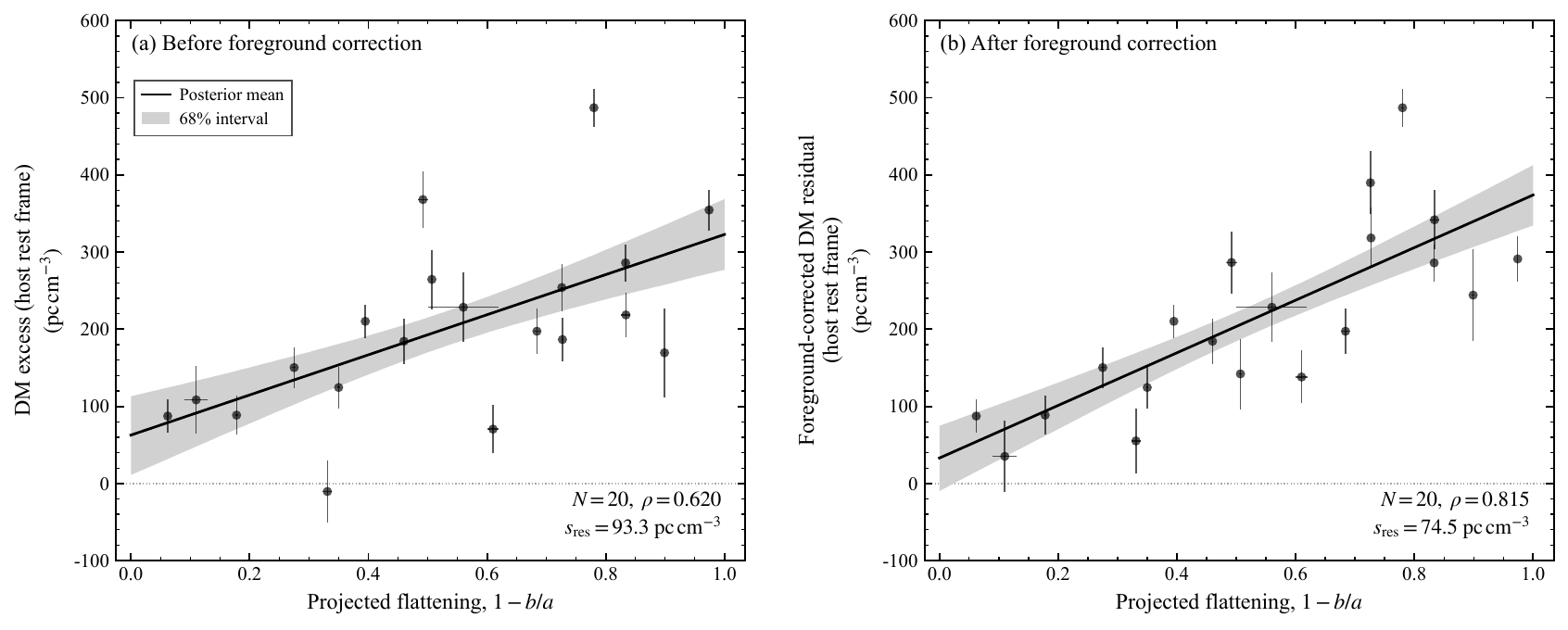}
\caption{Host shape and host-rest-frame DM residuals for the same 20 FRBs before (a) and after (b) foreground correction using the separate 57-FRB calibration. Each point represents one FRB. The horizontal coordinate is \(1-q\), where \(q=b/a\) is the projected optical minor-to-major axis ratio of the host galaxy. Lines and shading show the posterior-mean shape relations and their pointwise 68\% posterior intervals. Horizontal error bars propagate the reported statistical axis-ratio errors; vertical bars propagate Galactic-DM uncertainties and, in (b), the shared foreground-calibration uncertainty. Error bars span the 16th--84th percentiles. Annotations give Spearman \(\rho\) and the sample standard deviation of residuals about each fitted relation, \(s_{\rm res}\). The foreground-corrected residual in (b) may retain cosmic contributions not traced by the group catalog.}
\label{fig:host}
\end{figure*}

As shown in the left panel of Figure~\ref{fig:host}, the 20 hosts within the Gfinder footprint with optical axis-ratio measurements show larger host-rest-frame DM residuals at smaller axis ratios before foreground correction, consistent with the association reported by \citet{Marnoch2026arXiv261013062M}. The Spearman coefficient between \(1-q\) and the residual is \(\rho=0.620\), with a nominal two-sided \(p=3.58\times10^{-3}\). Accounting for foreground groups using the calibration from the other 57 FRBs strengthens this association to \(\rho=0.815\), with \(p=1.20\times10^{-5}\) (Figure~\ref{fig:host}b). Comparisons with the other host properties are summarized in Appendix~\ref{sec:appendix-host-calibration}.

The stronger rank association is accompanied by reduced scatter about the fitted shape relation. Including the foreground term lowers the host-rest-frame residual standard deviation from \(93.3\) to \(74.5~\mathrm{pc\,cm^{-3}}\), a reduction of \(20.2\%\). The resulting shape slope is \(A=340.9~\mathrm{pc\,cm^{-3}}\), with a 95\% posterior interval of \(190.5\)--\(488.8~\mathrm{pc\,cm^{-3}}\).

The gain extends to leave-one-out predictions of observer-frame \(DM_{\rm exgal}\). Relative to the axis-ratio model alone, including foreground information reduces the RMSE by \(16.2\%\) and the MAE by \(7.4\%\) (Table~\ref{tab:host-prediction-comparison}). In this 20-host subset, adding foreground information to a mean-host baseline does not improve predictions. Once host-shape variations are included, however, the same foreground calibration reduces the prediction errors, giving the lowest errors among the four models.

\begin{deluxetable}{lccc}
\tabletypesize{\footnotesize}
\tablewidth{\columnwidth}
\tablecaption{Scatter and leave-one-out prediction errors for the 20-host sample.\label{tab:host-prediction-comparison}}
\tablehead{
\colhead{Model} & \colhead{Residual SD} & \colhead{RMSE} & \colhead{MAE}
}
\startdata
Mean host & 115.7 & 96.7 & 72.8 \\
Foreground + mean host & 117.7 & 98.7 & 80.3 \\
Host shape & 93.3 & 76.3 & 54.2 \\
Foreground + host shape & 74.5 & 64.0 & 50.2
\enddata
\tablecomments{
Residual SD is the host-rest-frame in-sample standard deviation; RMSE and MAE are observer-frame leave-one-out prediction errors in \(DM_{\rm exgal}\). The 57-FRB calibration gives observer-frame \(k_{57}=49.6\,[29.8,69.5]\). The combined model gives host-rest-frame \(A=340.9\,[190.5,488.8]\), \(B=33.0\,[-54.7,118.5]\), and \(\sigma_{\rm H}=77.1\,[50.6,115.5]\). Here \(\sigma_{\rm H}\) is the fitted extra-scatter parameter, distinct from the empirical residual SD. Parameters are posterior means with equal-tailed 95\% intervals in brackets. All quantities are in \(\mathrm{pc\,cm^{-3}}\).
}
\end{deluxetable}

The stronger host-shape association and reduced residual scatter persist under the tested alternatives for Galactic subtraction and the cosmic mean. Across these variants, foreground correction reduces the residual standard deviation by \(19.5\%\)--\(22.9\%\), and the 95\% posterior interval for the shape slope remains above zero (Table~\ref{tab:robustness-summary}).

Using the mean random background and recalibrating the foreground term on the same 57 FRBs also strengthens the host-shape association, giving \(\rho=0.811\). The combined model yields a residual standard deviation of \(74.4~\mathrm{pc\,cm^{-3}}\) and leave-one-out RMSE and MAE of \(65.0\) and \(48.8~\mathrm{pc\,cm^{-3}}\), respectively, retaining the improvement over the shape-only model (Table~\ref{tab:host-prediction-sensitivity}). 

For the adopted median reference, the rank and scatter improvements persist in every single-source refit, and the RMSE and MAE gains remain positive after omitting each source's paired prediction errors.

\section{Discussion}
\label{sec:discussion}

\subsection{Foreground Group Environments and Ionized Gas}

The association between group-count excess and DM links electron-column variations to foreground group environments. The adopted \(1.5R_{180\mathrm{m}}\) aperture includes group interiors and outskirts, where intragroup gas, the circumgalactic media of member galaxies, and correlated intergalactic structure may all contribute. Stacked thermal Sunyaev--Zel'dovich measurements detect gas to approximately \(3R_{500}\) around nearby groups \citep{Pratt2021ApJ...920..104P}, while recent kinematic Sunyaev--Zel'dovich measurements provide complementary evidence for extended ionized gas around massive groups \citep{Gong2026arXiv260936355G}.

These observations motivate including group outskirts when tracing foreground gas. The counting aperture nevertheless defines an environmental scale, not a measured gas boundary. The intercepted electron column depends on halo mass, impact parameter, and gas distribution, whereas our count assigns equal weight to each intersected group. The fitted coefficient therefore describes an average DM dependence on the adopted environmental tracer; it does not assign the same physical electron column to every group or isolate the contribution of intragroup gas.

\subsection{Host Geometry and the Amplitude of DM Variations}

More inclined disks can present longer paths through ionized gas, providing a possible explanation for larger DM residuals toward smaller host axis ratios \citep{Xu2015RAA...15.1629X}. To assess whether this effect can reach the fitted amplitude, we used Milky Way electron-density models as idealized galaxy templates. We combined the large-scale NE2001 distribution \citep{Cordes2002astro.ph..7156C} with the YT2020 hot-gas model \citep{Yamasaki2020ApJ...888..105Y}, and repeated the calculation with NE2025 \citep{Ocker2026ApJ..1002....3O} replacing NE2001. For synthetic sources within each template, we varied the viewing inclination toward a distant external observer while keeping the gas distribution and source positions fixed.

As the optical axis ratio decreases from \(q=0.8\) to \(0.2\), our fitted relation implies a host-rest-frame mean DM increase of \(204.6~\mathrm{pc\,cm^{-3}}\), with a 95\% posterior interval of \(114.3\)--\(293.3~\mathrm{pc\,cm^{-3}}\). Over the same interval, the NE2001+YT2020 and NE2025+YT2020 templates give mean increases of \(181\)--\(240\) and \(158\)--\(203~\mathrm{pc\,cm^{-3}}\), respectively. These ranges span the adopted source distributions and optical geometries (Appendix~\ref{sec:appendix-inclination-models}). The comparable amplitudes support the physical plausibility of an inclination contribution under these assumptions. However, the templates are not matched to the gas densities, scale lengths, or burst locations of individual hosts, and optical axis ratio need not trace the geometry of their ionized gas.

\subsection{Foreground and Host Contributions to DM Scatter}

Accounting for both foreground groups and host shape yields the lowest residual scatter and leave-one-out prediction errors among the four models tested for the 20 hosts. Foreground information alone does not improve the mean-host baseline, suggesting that unmodeled host variation may obscure its predictive contribution. In this subset, more elongated hosts tend to lie along sightlines with smaller foreground group-count excesses (\(\rho=-0.297\), nominal two-sided \(p=0.203\)). This weak trend may contribute to the strengthening of the host-shape relation after foreground correction.

This empirical improvement does not uniquely partition the DM variance into foreground and host components. The corrected residuals still include cosmic structure untraced by the group catalog, and the optical measurements sample only part of the diversity of FRB hosts. Moreover, leave-one-out evaluation retains choices of observables and functional forms informed by the inspected sample. 

Larger, independently assembled FRB samples with more complete foreground catalogs and uniform host imaging can test whether these prediction gains persist across different foreground environments and host shapes. Combining FRB electron columns with thermal Sunyaev--Zel'dovich pressure measurements could constrain the density and thermal state of group-associated gas \citep{Munoz2018PhRvD..98j3518M}. Spatially resolved host spectroscopy and gas kinematics, together with burst offsets, could test whether optical shape traces the relevant ionized-gas path.

\section{Conclusions}
\label{sec:Conclusions}

In this work, we compiled a parent catalog of 136 localized FRBs with measured DMs and spectroscopic host redshifts, and examined foreground tracers and host-galaxy properties to investigate scatter about a fixed Macquart mean relation. The foreground group analysis uses 77 non-PRS FRBs within the Gfinder footprint. Of these, 20 have optical host axis-ratio measurements; the other 57 provide a disjoint foreground-calibration sample for the host analysis. Our main results are as follows:

\begin{enumerate}

\item The foreground group-count excess within \(1.5R_{180\mathrm{m}}\) is positively correlated with \(DM_{\rm excess}\), with a Spearman coefficient of \(\rho=0.459\). The fitted observer-frame coefficient is \(k=45.1~\mathrm{pc\,cm^{-3}}\) per unit group-count excess, with a 95\% posterior interval of \(27.8\)--\(62.3~\mathrm{pc\,cm^{-3}}\). Adding this term to the mean-host baseline reduces the leave-one-out RMSE and MAE in observer-frame \(DM_{\rm exgal}\) by \(13.0\%\) and \(7.5\%\), respectively.

\item For the 20-host sample, foreground correction using the separate 57-FRB calibration strengthens the correlation between the host-shape indicator \(1-q\) and host-rest-frame DM residuals: the Spearman coefficient increases from \(0.620\) to \(0.815\). Here, \(q=b/a\) is the optical minor-to-major axis ratio. Relative to the axis-ratio model alone, adding the foreground term reduces the host-rest-frame in-sample residual standard deviation by \(20.2\%\) and the observer-frame leave-one-out RMSE and MAE by \(16.2\%\) and \(7.4\%\), respectively.

\item As the optical axis ratio decreases from \(0.8\) to \(0.2\), the fitted host-shape relation implies a host-rest-frame mean DM increase of \(204.6~\mathrm{pc\,cm^{-3}}\), with a 95\% posterior interval of \(114.3\)--\(293.3~\mathrm{pc\,cm^{-3}}\). Models based on the Milky Way electron-density distribution produce inclination-induced DM changes of comparable amplitude, supporting the physical plausibility of an inclination contribution under the adopted assumptions.

\end{enumerate}

\begin{acknowledgments}
This work is supported by the National SKA Program of China No. 2025SKA0150103, National Natural Science Foundation of China under Nos. 12550002, 12133008, 12221003, 11890692. We acknowledge the science research grants from the China Manned Space Project with No. CMS-CSST-2021-A04 and No. CMS-CSST-2025-A10.
\end{acknowledgments}

\section*{Data and Code Availability}

The compiled FRB catalog, derived data products, and analysis code are available to the editor and referees on request. The original survey catalogs and published measurements are available from the sources cited in the text.

\bibliography{sample701}{}
\bibliographystyle{aasjournalv7}

\clearpage
\onecolumngrid
\appendix
\setcounter{figure}{0}
\renewcommand{\thefigure}{A\arabic{figure}}
\renewcommand{\theHfigure}{A\arabic{figure}}
\setcounter{table}{0}
\renewcommand{\thetable}{A\arabic{table}}
\renewcommand{\theHtable}{A\arabic{table}}
\setcounter{equation}{0}
\renewcommand{\theequation}{A\arabic{equation}}
\renewcommand{\theHequation}{A\arabic{equation}}

\section{Localized FRB Catalog}
\label{sec:appendix-frb-catalog}

\begin{figure}[htbp]
\centering
\includegraphics[width=\textwidth]{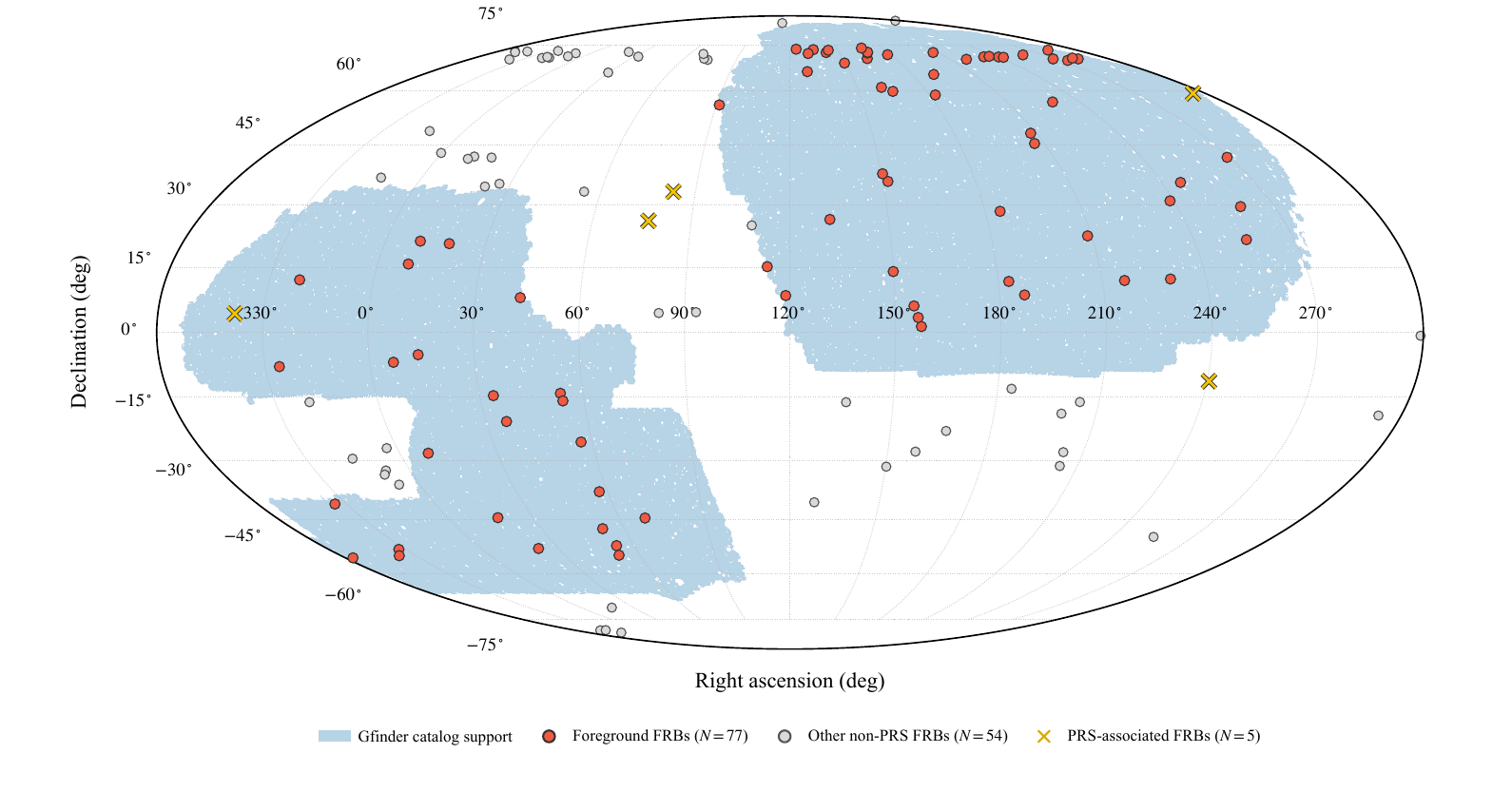}
\caption{Sky distribution of the 136 FRBs in the parent catalog. Shading shows the Gfinder catalog support used for sample selection, defined by occupied HEALPix pixels before the group-membership and redshift cuts. Red circles mark the 77-source foreground sample, gray circles show the other 54 non-PRS sources, and yellow crosses mark the five PRS-associated sources excluded from the analysis.}
\label{fig:sky-coverage}
\end{figure}

Table~\ref{tab:frb-sample} lists the parent catalog of 136 localized FRBs. Figure~\ref{fig:sky-coverage} shows their sky distribution together with the Gfinder catalog coverage. The optical axis ratios and DM quantities used for the 20-host analysis are listed separately in Table~\ref{tab:host-analysis-inputs}.

\begingroup
\setlength{\tabcolsep}{3pt}
\startlongtable
\begin{deluxetable}{lcccccccccc}
\returntotwocolfalse
\tabletypesize{\scriptsize}
\tablewidth{0pt}
\tablecaption{Parent catalog of 136 localized FRBs.\label{tab:frb-sample}}
\tablehead{
\colhead{FRB ID} &
\colhead{R.A.} &
\colhead{Decl.} &
\colhead{$\sigma_{\rm loc}$} &
\colhead{$DM_{\rm obs}$} &
\colhead{$z_{\rm host}$} &
\colhead{$DM_{\rm MW,disk}$\tablenotemark{a}} &
\colhead{$DM_{\rm MW,halo}$\tablenotemark{b}} &
\colhead{Gfinder} &
\colhead{PRS} &
\colhead{Loc.} \\
\colhead{} &
\colhead{(deg)} &
\colhead{(deg)} &
\colhead{(arcsec)} &
\colhead{($\mathrm{pc\,cm^{-3}}$)} &
\colhead{} &
\colhead{($\mathrm{pc\,cm^{-3}}$)} &
\colhead{($\mathrm{pc\,cm^{-3}}$)} &
\colhead{coverage} &
\colhead{excluded} &
\colhead{source}
}
\startdata
20121102A & 82.994589 & 33.147932 & 0.0043 & 558.10 & 0.19273 & 207.93 & 40.91 & No & Yes & 1 \\
20180301A & 93.226833 & 4.671056 & 0.619 & 536.00 & 0.33044 & 156.17 & 38.06 & No & No & 1 \\
20180916B & 29.503126 & 65.716754 & 0.0023 & 348.76 & 0.03370 & 125.65 & 43.08 & No & No & 1 \\
20180924B & 326.105229 & -40.900028 & 0.176 & 362.16 & 0.32120 & 39.79 & 45.54 & Yes & No & 1 \\
20181112A & 327.348458 & -52.970944 & 5.755 & 589.27 & 0.47550 & 39.47 & 45.04 & Yes & No & 1 \\
20181119A & 190.479946 & 65.111920 & 0.0100 & 366.10 & 0.26064 & 31.20 & 31.35 & Yes & No & 1 \\
20190102C & 322.415667 & -79.475694 & 0.959 & 364.55 & 0.29120 & 55.04 & 46.99 & No & No & 1 \\
20190303A & 207.995833 & 48.121111 & 12.000 & 222.40 & 0.06440 & 27.44 & 32.19 & Yes & No & 18 \\
20190303B & 128.687700 & 66.005900 & 11.000 & 193.43 & 0.06560 & 41.05 & 31.32 & Yes & No & 19 \\
20190417A & 294.774550 & 59.326897 & 0.0052 & 1378.90 & 0.12817 & 76.20 & 43.27 & Yes & Yes & 2 \\
20190425A & 255.675000 & 21.576389 & 12.000 & 128.20 & 0.03122 & 48.35 & 48.98 & Yes & No & 3 \\
20190502A & 165.086300 & 59.915600 & 16.000 & 625.74 & 0.28030 & 31.07 & 30.45 & Yes & No & 19 \\
20190520B & 240.517775 & -11.288147 & 0.100 & 1210.30 & 0.24100 & 60.37 & 69.16 & No & Yes & 1 \\
20190523A & 207.065000 & 72.469722 & 4.000 & 760.80 & 0.66000 & 34.99 & 32.39 & Yes & No & 1 \\
20190608B & 334.019875 & -7.898250 & 0.381 & 340.05 & 0.11778 & 36.10 & 38.89 & Yes & No & 1 \\
20190611B & 320.745583 & -79.397583 & 1.170 & 332.63 & 0.37780 & 55.44 & 47.21 & No & No & 1 \\
20190711A & 329.417225 & -80.358029 & 2.409 & 587.90 & 0.52172 & 54.05 & 46.20 & No & No & 1 \\
20190714A & 183.979708 & -13.021000 & 0.546 & 504.13 & 0.23650 & 35.73 & 36.44 & No & No & 1 \\
20191001A & 323.351696 & -54.748306 & 0.460 & 507.00 & 0.23400 & 41.84 & 46.71 & Yes & No & 1 \\
20191228A & 344.430529 & -29.594117 & 0.840 & 298.00 & 0.24300 & 30.23 & 36.88 & No & No & 1 \\
20200120E & 149.477833 & 68.818889 & 0.400 & 87.82 & 0.00080 & 36.35 & 31.10 & Yes & No & 1 \\
20200430A & 229.706454 & 12.376329 & 0.459 & 380.00 & 0.16100 & 29.54 & 42.24 & Yes & No & 1 \\
20200906A & 53.495546 & -14.083008 & 0.510 & 577.84 & 0.36880 & 32.66 & 30.16 & Yes & No & 1 \\
20201124A & 77.014615 & 26.060696 & 0.0040 & 411.00 & 0.09820 & 136.07 & 36.24 & No & Yes & 1 \\
20210117A & 339.979196 & -16.151445 & 0.418 & 728.95 & 0.21450 & 32.26 & 37.17 & No & No & 1 \\
20210320C & 204.458767 & -16.122670 & 0.471 & 384.59 & 0.27970 & 37.86 & 42.13 & No & No & 1 \\
20210405I & 255.339700 & -49.545150 & 0.500 & 565.17 & 0.06600 & 444.25 & 143.95 & No & No & 4 \\
20210410D & 326.086250 & -79.318194 & 0.800 & 575.00 & 0.14150 & 53.88 & 46.48 & No & No & 1 \\
20210603A & 10.274058 & 21.226270 & 1.080 & 500.15 & 0.17720 & 36.90 & 32.43 & Yes & No & 5 \\
20210807D & 299.221433 & -0.762361 & 0.500 & 251.30 & 0.12927 & 117.05 & 73.58 & No & No & 1 \\
20211127I & 199.808812 & -18.837911 & 0.510 & 234.97 & 0.04690 & 40.27 & 41.59 & No & No & 1 \\
20211203C & 204.562500 & -31.380267 & 0.440 & 635.00 & 0.34386 & 60.52 & 49.79 & No & No & 1 \\
20211212A & 157.350792 & 1.360431 & 0.500 & 209.00 & 0.07070 & 34.72 & 32.18 & Yes & No & 1 \\
20220105A & 208.803354 & 22.466228 & 1.513 & 580.00 & 0.27850 & 23.39 & 34.35 & Yes & No & 1 \\
20220204A & 274.226250 & 69.722503 & 1.560 & 612.20 & 0.40120 & 50.01 & 37.00 & Yes & No & 1 \\
20220207C & 310.199525 & 72.882327 & 0.630 & 262.30 & 0.04330 & 71.00 & 39.21 & No & No & 1 \\
20220208A & 322.575125 & 70.041042 & 0.990 & 437.00 & 0.35100 & 93.55 & 41.46 & No & No & 1 \\
20220222C & 203.904500 & -28.026925 & 0.550 & 1071.20 & 0.85300 & 53.19 & 47.40 & No & No & 6 \\
20220224C & 166.677542 & -22.939856 & 0.640 & 1140.20 & 0.62710 & 47.81 & 37.67 & No & No & 6 \\
20220307B & 350.874496 & 72.192386 & 1.720 & 499.15 & 0.24810 & 111.79 & 41.27 & No & No & 1 \\
20220310F & 134.720500 & 73.490833 & 1.900 & 462.15 & 0.47800 & 41.18 & 31.81 & Yes & No & 1 \\
20220319D & 32.177921 & 71.035260 & 0.580 & 110.95 & 0.01120 & 99.41 & 39.69 & No & No & 1 \\
20220330D & 163.751250 & 70.350750 & 1.620 & 468.10 & 0.37140 & 34.92 & 31.22 & Yes & No & 1 \\
20220418A & 219.105575 & 70.095944 & 1.356 & 623.45 & 0.62130 & 34.91 & 32.84 & Yes & No & 1 \\
20220501C & 352.379113 & -32.490731 & 0.440 & 449.50 & 0.38100 & 28.26 & 35.26 & No & No & 1 \\
20220506D & 318.044833 & 72.827278 & 1.110 & 396.93 & 0.30040 & 78.14 & 39.75 & No & No & 1 \\
20220509G & 282.670000 & 70.243833 & 2.200 & 269.50 & 0.08940 & 54.37 & 37.70 & Yes & No & 1 \\
20220529A & 19.104125 & 20.632375 & 1.390 & 246.40 & 0.18390 & 36.55 & 31.65 & Yes & No & 7 \\
20220610A & 351.073262 & -33.513850 & 0.445 & 1458.10 & 1.01500 & 28.58 & 35.64 & No & No & 1 \\
20220717A & 293.304167 & -19.287722 & 0.900 & 634.69 & 0.36295 & 106.08 & 86.47 & No & No & 8 \\
20220725A & 353.315204 & -35.990256 & 0.460 & 290.40 & 0.19260 & 28.54 & 35.42 & No & No & 1 \\
20220726A & 73.945542 & 69.929528 & 0.800 & 686.55 & 0.36190 & 81.18 & 35.32 & No & No & 1 \\
20220825A & 311.981450 & 72.584969 & 0.780 & 651.20 & 0.24140 & 73.12 & 39.45 & No & No & 1 \\
20220831A & 338.695542 & 70.538444 & 0.480 & 1146.25 & 0.26200 & 118.24 & 42.10 & No & No & 9 \\
20220912A & 347.270412 & 48.706641 & 0.0050 & 219.37 & 0.07710 & 115.68 & 43.44 & No & No & 10 \\
20220914A & 282.056792 & 73.336914 & 0.930 & 631.05 & 0.11390 & 52.95 & 36.95 & Yes & No & 1 \\
20220918A & 17.592142 & -70.811378 & 0.460 & 656.80 & 0.49100 & 39.04 & 38.66 & No & No & 1 \\
20220920A & 240.257083 & 70.918800 & 1.020 & 315.00 & 0.15820 & 38.70 & 34.12 & Yes & No & 1 \\
20221012A & 280.798708 & 70.524208 & 1.120 & 442.20 & 0.28470 & 53.22 & 37.46 & Yes & No & 1 \\
20221027A & 130.872000 & 72.100944 & 0.800 & 452.50 & 0.54220 & 41.74 & 31.76 & Yes & No & 1 \\
20221029A & 141.963417 & 72.452317 & 0.800 & 1391.05 & 0.97500 & 39.18 & 31.56 & Yes & No & 1 \\
20221101B & 342.215792 & 70.681500 & 1.200 & 490.70 & 0.23950 & 123.10 & 42.13 & No & No & 1 \\
20221106A & 56.704808 & -25.569810 & 0.598 & 343.80 & 0.20440 & 32.58 & 30.71 & Yes & No & 1 \\
20221113A & 71.411000 & 70.307389 & 1.400 & 411.40 & 0.25050 & 83.44 & 35.54 & No & No & 1 \\
20221116A & 21.210958 & 72.653750 & 1.000 & 640.60 & 0.27640 & 99.53 & 39.85 & No & No & 1 \\
20221219A & 257.629792 & 71.626842 & 1.540 & 706.71 & 0.55300 & 43.46 & 35.32 & Yes & No & 1 \\
20230124A & 231.917000 & 70.968086 & 1.400 & 590.00 & 0.09390 & 37.11 & 33.58 & Yes & No & 1 \\
20230125D & 150.205042 & -31.544658 & 0.260 & 640.08 & 0.32650 & 76.85 & 41.76 & No & No & 6 \\
20230203A & 151.661594 & 35.694096 & 17.978 & 417.30 & 0.14640 & 31.01 & 29.90 & Yes & No & 1 \\
20230216A & 156.472167 & 3.436825 & 1.200 & 828.29 & 0.53100 & 34.30 & 31.91 & Yes & No & 1 \\
20230222B & 238.739123 & 30.898695 & 11.498 & 188.03 & 0.11000 & 29.90 & 38.96 & Yes & No & 1 \\
20230307A & 177.781583 & 71.695131 & 0.200 & 610.15 & 0.27060 & 34.44 & 31.51 & Yes & No & 1 \\
20230311A & 91.109662 & 55.945953 & 15.739 & 364.34 & 0.19180 & 82.93 & 34.27 & Yes & No & 1 \\
20230501A & 340.027083 & 70.922139 & 0.900 & 533.70 & 0.30150 & 117.19 & 41.90 & No & No & 1 \\
20230506C & 12.099837 & 42.006090 & 0.180 & 772.00 & 0.38960 & 63.64 & 35.96 & No & No & 11 \\
20230521B & 351.036000 & 71.138031 & 0.559 & 1342.90 & 1.35400 & 120.81 & 41.95 & No & No & 9 \\
20230526A & 22.232617 & -52.717331 & 0.430 & 361.40 & 0.15700 & 30.37 & 34.06 & Yes & No & 1 \\
20230613A & 356.852708 & -27.052781 & 0.520 & 483.51 & 0.39230 & 27.56 & 33.95 & No & No & 6 \\
20230616A & 184.614579 & 28.356649 & 16.992 & 918.33 & 0.44850 & 21.55 & 31.13 & Yes & No & 1 \\
20230626A & 235.629583 & 71.133547 & 1.800 & 451.20 & 0.32700 & 37.83 & 33.81 & Yes & No & 1 \\
20230628A & 166.786708 & 72.281844 & 1.400 & 345.15 & 0.12700 & 35.51 & 31.44 & Yes & No & 1 \\
20230702A & 132.124693 & 26.389497 & 15.264 & 598.88 & 0.36580 & 41.54 & 30.81 & Yes & No & 1 \\
20230708A & 303.115542 & -55.356270 & 0.467 & 411.51 & 0.10500 & 57.77 & 58.03 & Yes & No & 1 \\
20230712A & 167.358529 & 72.557783 & 1.400 & 586.96 & 0.45250 & 35.61 & 31.47 & Yes & No & 1 \\
20230718A & 128.161929 & -40.451932 & 0.610 & 477.00 & 0.03500 & 490.24 & 56.40 & No & No & 1 \\
20230730A & 54.664564 & 33.159296 & 12.215 & 312.58 & 0.21150 & 75.33 & 33.75 & No & No & 1 \\
20230808F & 53.304125 & -51.935283 & 0.500 & 653.20 & 0.34720 & 34.11 & 33.38 & Yes & No & 12 \\
20230814B & 335.974750 & 73.025906 & 0.792 & 696.40 & 0.55350 & 96.64 & 40.53 & No & No & 9 \\
20230828A & 22.287034 & 34.429645 & 12.485 & 377.28 & 0.62680 & 49.70 & 33.39 & No & No & 1 \\
20230902A & 52.139775 & -47.333503 & 0.693 & 440.10 & 0.36190 & 32.82 & 32.63 & Yes & No & 1 \\
20230907D & 187.142500 & 8.658092 & 0.570 & 1030.79 & 0.46380 & 26.88 & 33.02 & Yes & No & 6 \\
20230918A & 161.050726 & 61.120861 & 11.617 & 365.96 & 0.06000 & 31.98 & 30.48 & Yes & No & 1 \\
20230923A & 206.614104 & 45.408008 & 13.813 & 918.33 & 0.44450 & 26.50 & 32.17 & Yes & No & 1 \\
20230926A & 269.124876 & 41.814304 & 17.964 & 222.81 & 0.05530 & 55.57 & 44.22 & Yes & No & 1 \\
20230930A & 10.507225 & 41.417337 & 0.180 & 456.00 & 0.09250 & 62.16 & 35.96 & No & No & 11 \\
20231005A & 246.028004 & 35.448712 & 11.390 & 188.84 & 0.07130 & 34.29 & 39.60 & Yes & No & 1 \\
20231006B & 109.619968 & 84.776136 & 16.042 & 1252.30 & 0.47800 & 49.60 & 33.67 & No & No & 1 \\
20231011A & 18.241101 & 41.749098 & 11.495 & 186.42 & 0.07830 & 63.55 & 35.36 & No & No & 1 \\
20231017A & 346.754287 & 36.652682 & 13.122 & 344.12 & 0.24500 & 62.44 & 39.10 & No & No & 1 \\
20231020B & 57.278208 & -37.769878 & 0.400 & 952.20 & 0.47750 & 32.94 & 31.66 & Yes & No & 6 \\
20231102A & 228.837416 & 56.799192 & 15.008 & 906.20 & 0.68780 & 32.41 & 33.76 & Yes & No & 1 \\
20231120A & 143.983958 & 73.284667 & 1.100 & 438.90 & 0.03680 & 39.14 & 31.62 & Yes & No & 1 \\
20231123A & 82.623249 & 4.475541 & 16.254 & 302.07 & 0.07290 & 79.56 & 34.22 & No & No & 1 \\
20231123B & 242.538167 & 70.785056 & 1.200 & 396.70 & 0.26210 & 39.14 & 34.28 & Yes & No & 1 \\
20231220A & 123.908708 & 73.659917 & 0.326 & 491.20 & 0.33550 & 44.16 & 32.12 & Yes & No & 9 \\
20231223C & 259.544649 & 29.497937 & 11.333 & 179.54 & 0.10590 & 48.83 & 46.34 & Yes & No & 1 \\
20231226A & 155.363733 & 6.110260 & 0.511 & 329.90 & 0.15690 & 33.82 & 31.60 & Yes & No & 1 \\
20231229A & 26.467825 & 35.112918 & 11.430 & 198.55 & 0.01900 & 51.85 & 33.35 & No & No & 1 \\
20231230D & 118.760958 & 8.514389 & 0.700 & 676.50 & 0.50600 & 68.91 & 34.34 & Yes & No & 20 \\
20240114A & 321.915979 & 4.329352 & 0.0930 & 527.65 & 0.13029 & 51.05 & 45.05 & Yes & Yes & 13 \\
20240117B & 53.877875 & -15.852111 & 1.290 & 881.00 & 0.64300 & 32.52 & 30.21 & Yes & No & 20 \\
20240123A & 68.262500 & 71.945283 & 0.606 & 1462.00 & 0.96800 & 82.11 & 35.65 & No & No & 9 \\
20240201A & 149.905608 & 14.087972 & 0.457 & 374.50 & 0.04273 & 33.78 & 30.88 & Yes & No & 1 \\
20240203C & 8.897542 & 15.840861 & 0.600 & 454.00 & 0.24260 & 34.11 & 32.12 & Yes & No & 20 \\
20240209A & 289.887500 & 86.064444 & 2.120 & 176.49 & 0.13840 & 51.43 & 34.92 & No & No & 14 \\
20240210A & 8.779575 & -28.270754 & 0.547 & 283.73 & 0.02369 & 26.38 & 32.38 & Yes & No & 1 \\
20240213A & 166.168292 & 74.075389 & 0.466 & 357.40 & 0.11850 & 36.49 & 31.63 & Yes & No & 9 \\
20240215A & 268.441250 & 70.232361 & 0.373 & 549.50 & 0.21000 & 47.31 & 36.38 & Yes & No & 9 \\
20240304A & 136.330825 & -16.166631 & 0.810 & 652.60 & 0.24230 & 65.09 & 37.09 & No & No & 1 \\
20240304B & 182.997042 & 11.813017 & 0.480 & 2458.20 & 2.14800 & 26.27 & 32.26 & Yes & No & 15 \\
20240310A & 17.621871 & -44.439424 & 0.588 & 601.80 & 0.12700 & 28.29 & 33.15 & Yes & No & 1 \\
20240312D & 51.623388 & -54.600972 & 1.290 & 333.80 & 0.04986 & 34.45 & 33.81 & Yes & No & 20 \\
20240318A & 150.393167 & 37.616361 & 0.790 & 256.40 & 0.11200 & 31.43 & 29.89 & Yes & No & 20 \\
20240525A & 6.782875 & -6.889611 & 0.480 & 491.60 & 0.32660 & 29.48 & 31.72 & Yes & No & 20 \\
20240615B & 33.963625 & -14.624500 & 0.660 & 201.00 & 0.07300 & 28.11 & 30.07 & Yes & No & 20 \\
20241027B & 36.029000 & -20.703528 & 0.640 & 320.00 & 0.33600 & 27.94 & 30.25 & Yes & No & 20 \\
20241228A & 216.385750 & 12.025042 & 10.930 & 246.53 & 0.16140 & 25.65 & 37.99 & Yes & No & 16 \\
20250206A & 338.670833 & 12.179167 & 1.880 & 207.12 & 0.11800 & 41.06 & 37.74 & Yes & No & 21 \\
20250227A & 113.377750 & 15.223269 & 0.450 & 156.60 & 0.03140 & 77.93 & 34.34 & Yes & No & 21 \\
20250316A & 182.434663 & 58.849086 & 0.0680 & 161.82 & 0.00650 & 29.26 & 30.78 & Yes & No & 17 \\
20250410E & 14.041667 & -5.138611 & 1.800 & 619.08 & 0.49800 & 28.10 & 30.99 & Yes & No & 21 \\
20250507A & 359.633333 & 42.942222 & 0.300 & 212.64 & 0.12040 & 69.51 & 37.99 & No & No & 21 \\
20250515A & 42.870833 & 8.036111 & 0.400 & 169.37 & 0.06250 & 35.65 & 30.19 & Yes & No & 21 \\
20250613A & 69.137667 & -44.532936 & 0.550 & 174.60 & 0.09870 & 38.95 & 33.14 & Yes & No & 22 \\
20250704A & 108.450000 & 24.960000 & 0.260 & 356.45 & 0.11300 & 81.86 & 34.00 & No & No & 21 \\
20251130A & 158.472708 & -27.891747 & 0.510 & 977.00 & 0.41910 & 59.71 & 39.47 & No & No & 23 \\
\enddata

\tablecomments{
The table lists FRBs with spectroscopic host-galaxy redshifts and
adopted localization scales $\sigma_{\rm loc}<20''$, where
$\sigma_{\rm loc}$ is the larger of the two adopted localization
error scales. The final column identifies the source of the
localization uncertainty.
For the DSA-110 entries with reference 9, the original localization
axes are from \citet{Connor2025NatAs...9.1226C}. We interpreted
these using the 90\% localization convention described by
\citet{Sharma2024Natur.635...61S}. Assuming joint bivariate-Gaussian
confidence regions, we divided both semi-axes by
$\sqrt{2\ln 10}=2.146$ to obtain the adopted $1\sigma$ axis scales
before taking $\sigma_{\rm loc}$.
}

\tablenotetext{a}{
Galactic interstellar DM calculated with NE2025 \citep{Ocker2026ApJ..1002....3O}.
}

\tablenotetext{b}{
Galactic hot-gas DM calculated with the flattened and spherical components of YT2020 \citep{Yamasaki2020ApJ...888..105Y}.
}

\tablerefs{
(1) \texttt{FRBs/FRB} compilation \citealt{Prochaska2025zndo..16763432P};
(2) \citealt{Moroianu2026ApJ...996L..16M};
(3) \citealt{Bhardwaj2024ApJ...971L..51B};
(4) \citealt{Driessen2024MNRAS.527.3659D};
(5) \citealt{Cassanelli2024NatAs...8.1429C};
(6) \citealt{Pastor2026MNRAS.545f2144P};
(7) \citealt{Pandhi2026ApJ..1000L..53P};
(8) \citealt{Rajwade2024MNRAS.532.3881R};
(9) \citealt{Connor2025NatAs...9.1226C};
(10) \citealt{Hewitt2024MNRAS.529.1814H};
(11) \citealt{Anna-Thomas2025ApJ...993..221A};
(12) \citealt{Hanmer2025MNRAS.538.1800H};
(13) \citealt{Bhardwaj2025ApJ...992L..35B};
(14) \citealt{Shah2025ApJ...979L..21S};
(15) \citealt{Caleb2025arXiv250801648C};
(16) \citealt{Curtin2026ApJ...998...97C};
(17) \citealt{CHIME2025ApJ...989L..48C};
(18) \citealt{Michilli2023ApJ...950..134M};
(19) \citealt{CHIME2024ApJ...969..145C};
(20) \citealt{Gordon2025ApJ...993..119G};
(21) \citealt{Dong2026arXiv260928629D};
(22) \citealt{Dial2026arXiv260700505D};
(23) \citealt{Tian2025ATel17548....1T}.
}
\end{deluxetable}
\endgroup

\begingroup
\renewcommand{\arraystretch}{1.2}
\begin{deluxetable}{lcrrrrr}
\tabletypesize{\small}
\tablewidth{0pt}
\tablecaption{Optical axis ratios and DM quantities for the 20-host analysis.\label{tab:host-analysis-inputs}}
\tablehead{
\colhead{FRB} &
\colhead{$q=b/a$} &
\colhead{$X$} &
\colhead{$DM_{\rm exgal}$} &
\colhead{$M(z)$} &
\colhead{$\overline{k}_{57}X$} &
\colhead{$DM_{\rm res}$}
}
\startdata
20180924B & $0.669\pm0.008$ & -1 & $276.8_{-30.1}^{+30.1}$ & 284.7 & $-49.6_{-10.0}^{+9.9}$ & $41.7_{-31.7}^{+31.8}$ \\
20181112A & $0.89\pm0.02$ & 1 & $504.8_{-29.8}^{+29.8}$ & 431.3 & $49.6_{-9.9}^{+10.0}$ & $23.8_{-31.5}^{+31.4}$ \\
20190608B & $0.540\pm0.002$ & 0 & $265.1_{-26.4}^{+26.4}$ & 100.1 & $0.0$ & $164.9_{-26.4}^{+26.4}$ \\
20191001A & $0.493\pm0.002$ & 2 & $418.4_{-31.2}^{+31.2}$ & 204.1 & $99.2_{-19.7}^{+20.1}$ & $115.1_{-37.2}^{+37.0}$ \\
20200430A & $0.316\pm0.006$ & 0 & $308.2_{-25.6}^{+25.6}$ & 138.3 & $0.0$ & $169.9_{-25.6}^{+25.6}$ \\
20200906A & $0.274\pm0.002$ & -2 & $515.0_{-22.1}^{+22.1}$ & 329.5 & $-99.2_{-20.1}^{+19.7}$ & $284.8_{-29.7}^{+29.9}$ \\
20211212A & $0.822\pm0.002$ & 0 & $142.1_{-23.5}^{+23.5}$ & 59.4 & $0.0$ & $82.7_{-23.5}^{+23.5}$ \\
20220105A & $0.0259\pm0.0009$ & 1 & $522.3_{-20.7}^{+20.7}$ & 245.0 & $49.6_{-9.9}^{+10.0}$ & $227.6_{-23.0}^{+22.9}$ \\
20221106A & $0.650\pm0.002$ & 0 & $280.5_{-22.3}^{+22.3}$ & 177.2 & $0.0$ & $103.3_{-22.3}^{+22.3}$ \\
20230708A & $0.44\pm0.06$ & 0 & $295.7_{-40.7}^{+40.7}$ & 89.0 & $0.0$ & $206.7_{-40.7}^{+40.7}$ \\
20230902A & $0.39\pm0.01$ & -1 & $374.6_{-23.0}^{+23.0}$ & 322.9 & $-49.6_{-10.0}^{+9.9}$ & $101.3_{-25.1}^{+25.2}$ \\
20231226A & $0.725\pm0.003$ & 0 & $264.5_{-23.0}^{+23.0}$ & 134.6 & $0.0$ & $129.8_{-23.0}^{+23.0}$ \\
20231230D & $0.101\pm0.002$ & -1 & $573.3_{-38.3}^{+38.3}$ & 460.8 & $-49.6_{-10.0}^{+9.9}$ & $162.1_{-39.5}^{+39.6}$ \\
20240117B & $0.508\pm0.009$ & 1 & $818.3_{-22.1}^{+22.1}$ & 594.3 & $49.6_{-9.9}^{+10.0}$ & $174.4_{-24.3}^{+24.2}$ \\
20240201A & $0.1666\pm0.0002$ & 0 & $309.8_{-22.8}^{+22.8}$ & 35.6 & $0.0$ & $274.2_{-22.8}^{+22.8}$ \\
20240203C & $0.166\pm0.008$ & -2 & $387.8_{-23.3}^{+23.3}$ & 212.0 & $-99.2_{-20.1}^{+19.7}$ & $275.0_{-30.6}^{+30.8}$ \\
20240210A & $0.6051\pm0.0005$ & 0 & $225.0_{-20.8}^{+20.8}$ & 19.7 & $0.0$ & $205.3_{-20.8}^{+20.8}$ \\
20240310A & $0.220\pm0.002$ & 0 & $540.4_{-21.7}^{+21.7}$ & 108.2 & $0.0$ & $432.1_{-21.7}^{+21.7}$ \\
20240525A & $0.273\pm0.001$ & -2 & $430.4_{-21.5}^{+21.5}$ & 289.8 & $-99.2_{-20.1}^{+19.7}$ & $239.9_{-29.3}^{+29.5}$ \\
20240615B & $0.938\pm0.001$ & 0 & $142.8_{-20.5}^{+20.5}$ & 61.4 & $0.0$ & $81.4_{-20.5}^{+20.5}$ \\
\enddata

\tablecomments{
Axis ratios and statistical fit errors are adopted from Table~1 of \citet{Marnoch2026arXiv261013062M}, which extends the measurements of \citet{Glowacki2025PASA...42..157G}. $X=N_{\rm obs}-\mathrm{median}(N_{\rm rand})$ is the direct group-count excess within $1.5R_{180\mathrm{m}}$. All DM quantities are in the observer frame and in $\mathrm{pc\,cm^{-3}}$. $DM_{\rm exgal}$ is the observed DM after Galactic subtraction, $M(z)$ is the fixed Macquart mean, and $DM_{\rm res}=DM_{\rm exgal}-M(z)-\overline{k}_{57}X$. The foreground coefficient is calibrated on the other 57 FRBs. DM errors span the 16th--84th percentiles, propagating the adopted Galactic uncertainties and, where applicable, the foreground-calibration uncertainty. The corrected residual can include cosmic contributions not traced by the group catalog.
}
\end{deluxetable}
\endgroup

\clearpage
\onecolumngrid
\setcounter{figure}{0}
\renewcommand{\thefigure}{B\arabic{figure}}
\renewcommand{\theHfigure}{B\arabic{figure}}
\setcounter{table}{0}
\renewcommand{\thetable}{B\arabic{table}}
\renewcommand{\theHtable}{B\arabic{table}}
\setcounter{equation}{0}
\renewcommand{\theequation}{B\arabic{equation}}
\renewcommand{\theHequation}{B\arabic{equation}}

\section{Foreground Tracers and Comparisons}
\label{sec:appendix-foreground-exploration}

We compared foreground counts and galaxy-density measures with observer-frame \(DM_{\rm excess}\), using the subset of the 131 non-PRS FRBs supported by each catalog. Table~\ref{tab:foreground-candidate-comparisons} summarizes representative results.

\subsection{Foreground Counts}

Gfinder coverage was defined by occupied HEALPix pixels at \(N_{\rm side}=1024\), using catalog entries before membership and redshift cuts \citep{Gorski2005ApJ...622..759G}. For each FRB, 10,000 group-control positions were sampled without replacement from occupied \(N_{\rm side}=128\) pixel centers within the coverage mask. Galaxy counts used 10,000 shared control positions sampled with replacement from the corresponding galaxy mask. Observed and random counts used identical foreground-redshift cuts and apertures. Covered sightlines without an eligible foreground interval were retained with zero observed and random counts.

For 2MRS, we adopted the published Galactic-latitude and extinction cuts \citep{Huchra2012ApJS..199...26H}. Cluster comparisons applied the PSZ2 survey mask to PSZ2, MCXC, and their combined catalog \citep{Planck2016A&A...594A..27P}.\footnote{PSZ2 catalog DOI: \url{https://doi.org/10.26131/IRSA449}; survey mask: \url{https://irsa.ipac.caltech.edu/data/Planck/release_2/catalogs/}.} Each mask used 10,000 shared random sightlines drawn uniformly in solid angle. Aperture centers were required to lie inside the mask, without a minimum aperture-area coverage requirement.

Since individual-galaxy halo radii were not available in the adopted tables, we used a common projected physical aperture of \(300~\mathrm{kpc}\) for the Gfinder and 2MRS galaxy-count comparisons. For 2MRS groups and the cluster catalogs, we used apertures of \(1\), \(1.5\), and \(2\) times their respective \(R_{200}\) and \(R_{500}\). We retained the published 2MRS group and MCXC radii and derived PSZ2 \(R_{500}\) from the catalog SZ mass proxy using Planck18. The combined catalog retained MCXC entries and added PSZ2 objects without an exact MCXC identifier match; its selection remains heterogeneous. Unless otherwise stated, count excesses subtract the median random count.

\subsection{Large-Scale Structure Tracers}

For DR1 LSS, we used BGS, LRG, and ELG samples over \(0.10\leq z<0.40\), \(0.40\leq z<1.10\), and \(1.10\leq z<1.60\), respectively, in shells of width 0.05 and \(20~\mathrm{Mpc}\) projected physical apertures. Shell contrasts were weighted galaxy counts divided by the expected counts from survey randoms, minus one. Catalog \texttt{WEIGHT} values were used, with random normalization performed separately by tracer, sky region, and shell.

WISE$\times$SuperCOSMOS used shell edges at \(z=0.005,0.05,0.10,\ldots,0.40\) and \(10\), \(20\), and \(30~\mathrm{Mpc}\) apertures. Shell contrasts were mask-corrected surface densities divided by the mean density in valid control apertures, minus one, using 10,000 random positions.

For both catalogs, angular apertures were evaluated at shell midpoints. Valid measurements required at least one complete foreground shell and support in every eligible shell within the catalog redshift range. Aperture centers had to lie inside the mask, with at least 80\% area coverage. DR1 LSS additionally required an effective random count of at least 25, defined as the expected count squared divided by its weighted random-sampling variance. Shell contrasts were summed with weights equal to the increase in the Macquart mean across each shell.

For 2M++, we used the galaxy-density reconstruction of \citet{Carrick2015MNRAS.450..317C}, smoothed on \(4\,h^{-1}\mathrm{Mpc}\). Using trilinear interpolation, we evaluated the local density at \(z=0.005\) for sources with \(z_{\rm FRB}>0.010\), and a column measure proportional to \(\int\delta_g(r)[1+z(r)]\,\mathrm{d}r\), where \(r\) is comoving distance. Column integrals began at \(z=0.005\) and retained the 0.05 foreground-redshift buffer. We limited the integration to \(113\,h^{-1}\mathrm{Mpc}\), leaving three smoothing lengths inside our adopted common radial limit of \(125\,h^{-1}\mathrm{Mpc}\), and excluded sightlines through the Galactic zone of avoidance, where missing galaxy data were filled using neighboring regions.

\subsection{Comparison Results}

The Gfinder galaxy-count excess within \(300~\mathrm{kpc}\) is positively associated with \(DM_{\rm excess}\), with \(\rho=0.391\), compared with \(\rho=0.459\) for the \(1.5R_{180\mathrm{m}}\) group-count excess on the same 77 FRBs. The positive galaxy-count association is qualitatively consistent with the correlation between excess cosmic DM and foreground galaxy overdensity reported by \citet{Hussaini2025ApJ...993L..27H}.

The 2MRS group counts also show a positive association, with \(\rho=0.289\) at \(1.5R_{200}\). Nonzero observed group counts occur along eight of its 128 supported sightlines, compared with 48 of the 77 Gfinder sightlines at \(1.5R_{180\mathrm{m}}\). Whereas 2MRS primarily probes nearby structure, Gfinder extends to \(z\sim1\), tracing group environments over a longer foreground path \citep{Huchra2012ApJS..199...26H,Yang2021ApJ...909..143Y,Wang2022ApJ...936..161W}. This greater depth supports its use in the host analysis.

Cluster intersections are sparse: only two of the 117 supported sightlines intersect MCXC objects within \(1\)--\(2R_{500}\), and none intersects a PSZ2 object. The DR1 LSS density measurements satisfy the angular and shell-support requirements for only ten FRBs. Because the catalog samples differ in coverage and membership, their correlations do not establish a common-sample ranking of the tracers.

\startlongtable
\begin{deluxetable}{lccc}
\tabletypesize{\footnotesize}
\tablewidth{0pt}
\tablecaption{Representative foreground-tracer correlations with observer-frame \(DM_{\rm excess}\).\label{tab:foreground-candidate-comparisons}}
\tablehead{
\colhead{Tracer} &
\colhead{\(N\)} &
\colhead{\(\rho\)} &
\colhead{\(p\)}
}
\startdata
\multicolumn{4}{l}{\textbf{Gfinder}}\\
Group-count excess, \(1.5R_{180\mathrm{m}}\) & 77 & 0.459 & \(2.72\times10^{-5}\)\\
Galaxy-count excess, \(300~\mathrm{kpc}\) & 77 & 0.391 & \(4.32\times10^{-4}\)\\[5pt]
\multicolumn{4}{l}{\textbf{2MRS}}\\
Galaxy-count excess, \(300~\mathrm{kpc}\) & 128 & 0.217 & 0.014\\
Group-count excess, \(1.5R_{200}\) & 128 & 0.289 & \(9.33\times10^{-4}\)\\[5pt]
\multicolumn{4}{l}{\textbf{MCXC and MCXC--PSZ2}}\\
Cluster-count excess, \(1.5R_{500}\) & 117 & 0.217 & 0.019\\[5pt]
\multicolumn{4}{l}{\textbf{PSZ2}}\\
Cluster-count excess, \(1.5R_{500}\) & 117 & --- & ---\\[5pt]
\multicolumn{4}{l}{\textbf{DR1 LSS}}\\
Macquart-weighted contrast, \(20~\mathrm{Mpc}\) & 10 & 0.479 & 0.162\\[5pt]
\multicolumn{4}{l}{\textbf{WISE$\times$SuperCOSMOS}}\\
Macquart-weighted contrast, \(10~\mathrm{Mpc}\) & 74 & 0.260 & 0.026\\
Macquart-weighted contrast, \(20~\mathrm{Mpc}\) & 70 & 0.126 & 0.300\\[5pt]
\multicolumn{4}{l}{\textbf{2M++}}\\
Local density at \(z=0.005\) & 129 & 0.306 & \(4.26\times10^{-4}\)\\
Signed column, \(0.05\) redshift buffer & 116 & 0.139 & 0.136
\enddata
\tablecomments{\(N\) is the number of FRBs with valid measurements; catalog samples differ in coverage and membership. The \(p\) values are nominal, two-sided asymptotic Spearman probabilities without multiple-testing adjustment. For PSZ2, the count excess is constant across the 117 supported sightlines, so \(\rho\) and \(p\) are undefined and shown as dashes. MCXC and the combined catalog give identical count excesses in the listed comparison.}
\end{deluxetable}

We compared galaxy-count, group-count, and joint models on the same 77 FRBs, using the \(300~\mathrm{kpc}\) galaxy-count excess and the \(1.5R_{180\mathrm{m}}\) group-count excess. The joint model extends Equation~\ref{eq:methods-main-regression} with separate galaxy and group coefficients, refitted in every leave-one-out fold. In this fixed-aperture comparison, adding group counts lowers both prediction errors relative to galaxy counts alone. Adding galaxy counts to the group model slightly reduces the in-sample scatter but increases both prediction errors (Table~\ref{tab:galaxy-group-comparison}). These results support our use of the group-count excess as the primary foreground tracer.

\begin{deluxetable}{lccc}
\tabletypesize{\footnotesize}
\tablewidth{0pt}
\tablecaption{Galaxy- and group-count model comparison for the same 77 FRBs.\label{tab:galaxy-group-comparison}}
\tablehead{
\colhead{Model} &
\colhead{Residual SD} &
\colhead{RMSE} &
\colhead{MAE}
}
\startdata
Mean host & 146.2 & 147.1 & 115.1\\
Mean host + galaxy counts & 132.4 & 136.8 & 111.8\\
Mean host + group counts & 125.2 & 128.0 & 106.4\\
Mean host + galaxy and group counts & 124.5 & 129.9 & 107.1
\enddata
\tablecomments{All models include the fixed cosmic mean and a fitted mean-host term. Galaxy and group counts use \(300~\mathrm{kpc}\) and \(1.5R_{180\mathrm{m}}\) apertures, respectively, with median random backgrounds. Residual SD measures in-sample scatter about the full-sample posterior-mean relation; RMSE and MAE measure leave-one-out prediction errors in \(DM_{\rm exgal}\). All quantities are in the observer frame and in \(\mathrm{pc\,cm^{-3}}\).}
\end{deluxetable}
\onecolumngrid

\clearpage
\onecolumngrid
\setcounter{figure}{0}
\renewcommand{\thefigure}{C\arabic{figure}}
\renewcommand{\theHfigure}{C\arabic{figure}}
\setcounter{table}{0}
\renewcommand{\thetable}{C\arabic{table}}
\renewcommand{\theHtable}{C\arabic{table}}
\setcounter{equation}{0}
\renewcommand{\theequation}{C\arabic{equation}}
\renewcommand{\theHequation}{C\arabic{equation}}

\section{Additional Host Observables}

\subsection{Derived Quantities and Measurement Selection}
\label{sec:appendix-derived-host-quantities}

For the surface-density comparisons, we normalized stellar mass and SFR by \(2\pi R_e^2\), using effective radii from single-component fits. The S\'ersic-index comparison included only freely fitted indices from single-component models. Normalized projected offsets were \(R_{\rm offset}/R_e\), using the published disk effective radius for multicomponent hosts.

For the gas-phase metallicities from \citet{Yamasaki2026ApJ...997L...6Y}, we retained hosts satisfying their emission-line selection criteria. For the measurements from \citet{Muller2026ApJ..1001..118M}, we excluded the source classified as an active galactic nucleus and values reported only as limits or bounded ranges, and analyzed the N2 and O3N2 metallicity estimates separately. We retained the tabulated O3N2 estimate for FRB20231230D, although its underlying H\(\beta\) measurement is an upper limit, and tested its exclusion separately.

The inclination-to-axis-ratio conversion for the AutoProf and Photutils measurements was
\begin{equation}
q=\left(\cos^2 i+0.2^2\sin^2 i\right)^{1/2},
\end{equation}
where \(i\) is the reported photometric inclination.

\subsection{Host-Property Correlations}
\label{sec:appendix-host-calibration}

None of the additional host-property comparisons in Table~\ref{tab:host-calibration-sensitivity} gives a nominal \(p<0.05\). Recalibrating the foreground after excluding the corresponding host subset leaves this conclusion unchanged.

\begin{deluxetable}{lccc}
\tabletypesize{\scriptsize}
\tablewidth{0pt}
\tablecaption{Additional host-property correlations with foreground-corrected host-rest-frame DM residuals.\label{tab:host-calibration-sensitivity}}
\tablehead{
\colhead{Observable} &
\colhead{\(N\)} &
\colhead{\(\rho\)} &
\colhead{\(p\)}
}
\startdata
\multicolumn{4}{l}{\textbf{Stellar populations}}\\[2pt]
Stellar mass & 27 & \(-0.138\) & 0.493\\
SFR averaged over 100 Myr & 27 & 0.111 & 0.581\\
Specific SFR & 27 & 0.302 & 0.126\\
Mass-weighted stellar age & 27 & 0.278 & 0.160\\
Stellar metallicity & 27 & \(-0.082\) & 0.686\\[6pt]
\multicolumn{4}{l}{\textbf{Galaxy structure and FRB location}}\\[2pt]
Single-component effective radius & 37 & \(-0.127\) & 0.452\\
S\'ersic index & 15 & \(-0.465\) & 0.081\\
Projected FRB offset & 43 & \(-0.177\) & 0.255\\
Normalized FRB offset & 43 & \(-0.090\) & 0.564\\[6pt]
\multicolumn{4}{l}{\textbf{Surface-density proxies}}\\[2pt]
Stellar-mass surface density & 24 & 0.037 & 0.865\\
SFR surface density & 24 & 0.285 & 0.177\\[6pt]
\multicolumn{4}{l}{\textbf{Emission-line measurements}}\\[2pt]
Gas-phase metallicity (Yamasaki et al.) & 17 & 0.375 & 0.138\\
H\(\alpha\)-based SFR (Muller et al.) & 7 & \(-0.090\) & 0.848\\
Gas-phase metallicity (Muller et al., N2) & 2 & --- & ---\\
Gas-phase metallicity (Muller et al., O3N2) & 4 & 0.316 & 0.684\\[6pt]
\multicolumn{4}{l}{\textbf{Alternative axis-ratio measurements, \(1-q\)}}\\[2pt]
AutoProf & 10 & 0.407 & 0.243\\
Photutils & 10 & 0.366 & 0.298
\enddata
\tablecomments{\(N\) is the number of hosts with the required measurements. Statistics use the adopted measurement values and the posterior-mean 77-FRB foreground calibration. The \(p\) values are nominal, two-sided asymptotic Spearman probabilities. Dashes indicate \(N<3\).}
\end{deluxetable}

Excluding the three sources identified in Appendix~\ref{sec:appendix-robustness} from the stellar-population comparisons leaves 24 hosts. Mass-weighted stellar age then shows a tentative positive correlation with the foreground-corrected DM residuals (\(\rho=0.413\), nominal \(p=0.045\)), using the other 53 FRBs in the original 77-FRB sample, including these three sources, for foreground calibration. Excluding FRB20231230D from the O3N2 comparison leaves three hosts, with \(\rho=0.866\) and nominal \(p=0.333\) under either calibration. Differences in sample membership and measurement definitions limit direct comparison with the GALFIT relation, and these separate rank tests do not establish whether host shape provides information beyond stellar age or the other properties.

\clearpage
\onecolumngrid
\setcounter{figure}{0}
\renewcommand{\thefigure}{D\arabic{figure}}
\renewcommand{\theHfigure}{D\arabic{figure}}
\setcounter{table}{0}
\renewcommand{\thetable}{D\arabic{table}}
\renewcommand{\theHtable}{D\arabic{table}}
\setcounter{equation}{0}
\renewcommand{\theequation}{D\arabic{equation}}
\renewcommand{\theHequation}{D\arabic{equation}}

\section{Robustness Tests}
\label{sec:appendix-robustness}

\subsection{Foreground Relation}

The 95\% posterior interval for \(k\) remains above zero under the tested changes to Galactic subtraction, the cosmic mean, counting aperture, and random-background reference, including imaging-matched controls, and when restricting the sample to \(z\leq1\) (Table~\ref{tab:robustness-summary}a).

To test sensitivity to spatial variations in imaging conditions, we matched random sightlines to each FRB using four DR9 brick-level quantities: \(z\)-band galaxy detection depth, \(z\)-band PSF size, Galactic reddening \(E(B-V)\), and the natural logarithm of one plus the point-source density in \(\mathrm{deg}^{-2}\) \citep{Dey2019AJ....157..168D}. Candidate positions satisfied the Gfinder coverage requirement and had valid measurements from survey-primary bricks with positive \(z\)-band exposure counts. Within the northern and southern regions separately, each quantity was centered on its candidate median and divided by \(1.4826\) times its median absolute deviation, using the standard deviation if this scale was zero. For each FRB, we drew 10,000 controls with replacement from the 2,048 nearest candidates by Euclidean distance in this standardized space. We recalculated the median random background and refitted the foreground model, keeping the observed counts unchanged. With these imaging-matched controls, the leave-one-out RMSE and MAE are \(129.6\) and \(107.3~\mathrm{pc\,cm^{-3}}\), respectively, both below those of the mean-host baseline.

Excluding FRB20220330D, FRB20221027A, and FRB20230216A, whose host associations were identified as uncertain by \citet{Sharma2024Natur.635...61S}, leaves 74 FRBs in the foreground-analysis sample. The positive foreground relation persists (Table~\ref{tab:robustness-summary}a). Adding the group-count term reduces the leave-one-out RMSE from \(146.1\) to \(128.0~\mathrm{pc\,cm^{-3}}\) and the MAE from \(113.8\) to \(106.0~\mathrm{pc\,cm^{-3}}\).

We propagated localization uncertainties through 4,096 position realizations, recomputing group counts and refitting the foreground model while holding sample membership and random backgrounds fixed. We used elliptical errors where both axes and the position angle were available, and circular errors based on the largest adopted localization scale otherwise. The foreground association changes little under these perturbations or the removal of individual FRBs (Table~\ref{tab:robustness-summary}c).

\begin{deluxetable}{lccc}
\tabletypesize{\footnotesize}
\tablewidth{0pt}
\tablecaption{Sensitivity of the foreground and host fits.\label{tab:robustness-summary}}
\tablehead{
\multicolumn{4}{c}{\textbf{(a) Foreground fits}}\\
\colhead{Configuration} & \colhead{\(N\)} & \colhead{\(\rho\)} & \colhead{\(k\) (95\% interval)}
}
\startdata
Main analysis & 77 & 0.459 & \(45.1\,[27.8,62.3]\)\\
NE2001 disk & 77 & 0.460 & \(45.2\,[27.9,62.5]\)\\
YMW16 disk & 77 & 0.469 & \(45.5\,[28.1,62.9]\)\\
Constant halo DM of 30 & 77 & 0.456 & \(45.2\,[28.0,62.3]\)\\
Non-diffuse fraction \(\times0.7\) & 77 & 0.450 & \(43.3\,[26.4,60.1]\)\\
Non-diffuse fraction \(\times1.3\) & 77 & 0.452 & \(46.9\,[29.1,64.7]\)\\
Mean random background & 77 & 0.411 & \(44.2\,[26.4,61.9]\)\\
Imaging-matched random sightlines & 77 & 0.450 & \(44.1\,[26.2,62.0]\)\\
\(z\leq1\) & 76 & 0.440 & \(42.8\,[25.3,60.4]\)\\
Excluding uncertain host associations & 74 & 0.435 & \(44.1\,[26.3,61.8]\)\\
\(R_{180\mathrm{m}}\) aperture & 77 & 0.392 & \(70.5\,[43.1,97.9]\)\\
\(2R_{180\mathrm{m}}\) aperture & 77 & 0.382 & \(30.2\,[19.3,41.2]\)\\
\tableline
\multicolumn{4}{c}{\textbf{(b) Propagation of DM-baseline changes to the 20-host fits}}\\[3pt]
Configuration & \(A\) (95\% interval) & Shape residual SD & Combined residual SD\\
\tableline
Main analysis & \(340.9\,[190.5,488.8]\) & 93.3 & 74.5\\
NE2001 disk & \(342.1\,[191.9,489.9]\) & 94.0 & 74.7\\
YMW16 disk & \(335.1\,[183.9,482.6]\) & 95.8 & 76.5\\
Constant halo DM of 30 & \(337.7\,[193.0,479.8]\) & 92.7 & 71.5\\
Non-diffuse fraction \(\times0.7\) & \(331.3\,[178.6,480.7]\) & 94.6 & 76.2\\
Non-diffuse fraction \(\times1.3\) & \(350.3\,[198.8,500.6]\) & 93.5 & 74.7\\
\tableline
\multicolumn{4}{c}{\textbf{(c) Foreground localization and individual-source influence}}\\[3pt]
Test & \(N\) & Range of \(\rho\) & Range of \(\overline{k}\)\\
\tableline
Localization uncertainty & 77 & \(0.447\)--\(0.473\) & \(43.2\)--\(45.6\)\\
Single-source deletion & 76 & \(0.436\)--\(0.498\) & \(40.3\)--\(48.0\)
\enddata
\tablecomments{Each row changes only the indicated setting. Coefficient estimates in panels (a) and (b) are posterior means with equal-tailed 95\% intervals. The foreground coefficient \(k\) is in the observer frame; \(A\) and the residual standard deviations are in the host rest frame. All DM quantities are in \(\mathrm{pc\,cm^{-3}}\). The non-diffuse fraction is \(1-f_{\rm d}(z)\). Panel (b) reports full-sample fits, with \(A\) for the combined model and the foreground recalibrated on the other 57 FRBs for each DM baseline. Shape and Combined denote the shape-only and foreground-plus-shape models, respectively. In (c), \(\overline{k}\) denotes the posterior mean from each fit. Localization ranges are the 2.5th--97.5th percentiles across position realizations; deletion ranges are minima and maxima across the reduced samples. These ranges are not posterior intervals.}
\end{deluxetable}
\onecolumngrid

\subsection{Host Relation and Predictions}

Propagating the Galactic and cosmic DM alternatives through the 57-FRB calibration to the 20-host fits preserves the stronger rank association and lower scatter after foreground correction. The residual standard deviation decreases by \(19.5\%\)--\(22.9\%\), and the 95\% posterior interval for \(A\) remains above zero (Table~\ref{tab:robustness-summary}b).

Removing these three sources from the foreground calibration leaves 54 calibrators and all 20 axis-ratio hosts unchanged. The corrected rank coefficient remains \(\rho=0.815\), and the combined model retains lower residual scatter and prediction errors than the shape-only model (Table~\ref{tab:host-prediction-sensitivity}).

\citet{Marnoch2026arXiv261013062M} identified the single-S\'ersic fits for FRB20190608B and FRB20231226A as primarily tracing their hosts' bulges rather than their disks. Excluding these two hosts and retaining the original 57-FRB calibration leaves 18 hosts for which foreground correction increases \(\rho\) from \(0.562\) to \(0.806\) and reduces both the fitted residual scatter and the leave-one-out prediction errors (Table~\ref{tab:host-prediction-sensitivity}).

After omitting each of the 20 hosts in turn, we refitted the shape-only and combined models with the same 57-FRB calibration. Foreground correction strengthens the rank association in every remaining sample and lowers the residual standard deviation by \(12.1\)--\(24.4~\mathrm{pc\,cm^{-3}}\). Removing each host's paired errors from the existing leave-one-out predictions, without refitting, leaves RMSE and MAE reductions of \(9.0\)--\(15.7\) and \(1.1\)--\(8.7~\mathrm{pc\,cm^{-3}}\), respectively.

For the ten Gfinder-covered hosts with AutoProf and Photutils measurements from \citet{Bhardwaj2024Natur.634.1065B}, we calibrated the foreground on the other 67 FRBs. We propagated inclination rounding through uniform perturbations of \(\pm0.5^\circ\) before conversion to axis ratios. For both measurement sets, the combined foreground-plus-shape model gives the lowest in-sample residual standard deviation among the compared models: \(107.8\) and \(110.0~\mathrm{pc\,cm^{-3}}\) for AutoProf and Photutils, respectively. Foreground correction also lowers the leave-one-out errors relative to the shape-only models, but the foreground-plus-mean-host model gives the lowest prediction errors (Table~\ref{tab:host-prediction-sensitivity}). The 95\% shape-slope intervals include zero. An assumed additional Gaussian inclination uncertainty with standard deviation \(1^\circ\) gives the same qualitative result. Differences in imaging and shape measurements between the studies \citep{Bhardwaj2024Natur.634.1065B,Marnoch2026arXiv261013062M}, together with differences in sample membership, may contribute to the weaker associations in these subsets.

\begin{table*}[t]
\centering
\caption{Host-model sensitivity to the background reference, sample, and shape measurements.}
\label{tab:host-prediction-sensitivity}
\small
\renewcommand{\arraystretch}{1.15}
\begin{tabular*}{\textwidth}{@{\extracolsep{\fill}}lccc@{}}
\toprule
Model & Residual SD & RMSE & MAE\\
\midrule
\multicolumn{4}{l}{\textbf{Mean background: 20 hosts, 57 calibrators}}\\[2pt]
Host shape & 93.3 & 76.3 & 54.2\\
Foreground + host shape & 74.4 & 65.0 & 48.8\\[6pt]
\multicolumn{4}{l}{\textbf{Uncertain associations excluded: 20 hosts, 54 calibrators}}\\[2pt]
Host shape & 93.3 & 76.3 & 54.2\\
Foreground + host shape & 74.4 & 63.9 & 50.0\\[6pt]
\multicolumn{4}{l}{\textbf{Excluding bulge-dominated fits: 18 hosts, 57 calibrators}}\\[2pt]
Host shape & 98.5 & 80.8 & 59.9\\
Foreground + host shape & 78.4 & 67.6 & 54.5\\[6pt]
\multicolumn{4}{l}{\textbf{AutoProf: 10 hosts, 67 calibrators}}\\[2pt]
Foreground + mean host & 121.1 & 104.9 & 82.0\\
Host shape & 146.9 & 134.5 & 97.9\\
Foreground + host shape & 107.8 & 109.5 & 86.6\\[6pt]
\multicolumn{4}{l}{\textbf{Photutils: 10 hosts, 67 calibrators}}\\[2pt]
Foreground + mean host & 121.1 & 104.9 & 82.0\\
Host shape & 145.8 & 132.2 & 97.3\\
Foreground + host shape & 110.0 & 107.2 & 87.6\\
\bottomrule
\end{tabular*}
\par\smallskip
\begin{minipage}{\textwidth}
\footnotesize\raggedright
\textit{Note.}---Residual SD is the host-rest-frame in-sample standard deviation; RMSE and MAE are observer-frame leave-one-out prediction errors in \(DM_{\rm exgal}\). All values are in \(\mathrm{pc\,cm^{-3}}\). Each block compares models fitted to the same hosts, with the stated disjoint foreground-calibration sample. For the uncertain-association test, only the foreground-calibration sample is reduced; all 20 hosts are retained. All blocks except the first use the median random background. AutoProf and Photutils use the same ten hosts.
\end{minipage}
\end{table*}

\clearpage
\onecolumngrid
\setcounter{figure}{0}
\renewcommand{\thefigure}{E\arabic{figure}}
\renewcommand{\theHfigure}{E\arabic{figure}}
\setcounter{table}{0}
\renewcommand{\thetable}{E\arabic{table}}
\renewcommand{\theHtable}{E\arabic{table}}
\setcounter{equation}{0}
\renewcommand{\theequation}{E\arabic{equation}}
\renewcommand{\theHequation}{E\arabic{equation}}

\section{Inclination-dependent DM Models}
\label{sec:appendix-inclination-models}

Following \citet{Xu2015RAA...15.1629X}, we used Milky Way electron-density models as idealized galaxy templates to assess the amplitude of an inclination effect. We retained the thick disk, thin disk, spiral arms, and Galactic center of NE2001 \citep{Cordes2002astro.ph..7156C}, excluding Solar-neighborhood structures, individual clumps, and voids, and used the corresponding NE2025 components for comparison \citep{Ocker2026ApJ..1002....3O}. Both templates included the flattened and spherical YT2020 components to represent the model galaxy's hot gaseous halo, with gas metallicity \(Z_{\rm gas}=0.3Z_\odot\) \citep{Yamasaki2020ApJ...888..105Y}. All components shared the same galactic center, and the disk planes were aligned. We kept the gas-density parameters at their published values. Possible overlap between the NE and YT2020 gas components has not been quantified \citep{Yamasaki2020ApJ...888..105Y,Price2021PASA...38...38P}.

We drew 10,000 source positions from the Gaussian radial surface-density and exponential vertical profiles of \citet{Xu2015RAA...15.1629X}, with uniform azimuths and probability densities
\begin{equation}
p(R)\propto R\exp\!\left(-\frac{R^2}{2\sigma_R^2}\right),
\qquad
p(Z)\propto\exp\!\left(-\frac{|Z|}{h_{\rm src}}\right),
\label{eq:appendix-source-distribution}
\end{equation}
where \(R\) and \(Z\) are the cylindrical galactocentric radius and height relative to the disk mid-plane, respectively. We adopted \(\sigma_R=6.5~\mathrm{kpc}\) and \(h_{\rm src}=330~\mathrm{pc}\), restricting sources to \((R^2+Z^2)^{1/2}<15~\mathrm{kpc}\). For the thinner \(h_{\rm src}=100~\mathrm{pc}\) population considered in that study, we retained the radial and azimuthal coordinates and matched vertical quantiles.

For each source at \(\boldsymbol r_{\rm FRB}\), we integrated toward a distant observer along the common direction \(\hat{\boldsymbol n}(i)=(\sin i,0,\cos i)\), where \(i\) is measured from the disk normal and the viewing azimuth is zero:
\begin{equation}
DM(i)=\sum_j\int_0^{s_{\max,j}}n_{e,j}\!\left[\boldsymbol r_{\rm FRB}+s\hat{\boldsymbol n}(i)\right]\,\mathrm{d}s.
\label{eq:appendix-inclination-dm}
\end{equation}
Here \(s\) is the path length in pc, and the sum runs over the gas components. Each integral ends at the forward intersection with a galactocentric sphere of radius \(15~\mathrm{kpc}\) for the NE distribution or \(260~\mathrm{kpc}\) for YT2020. Source positions and gas distributions remained fixed as \(i\) varied, and both templates used the same sources and viewing directions.

For an oblate optical disk aligned with the gas plane, the projected minor-to-major axis ratio \(q=b/a\) satisfies
\begin{equation}
q^2=\cos^2 i+q_0^2\sin^2 i.
\label{eq:appendix-axis-inclination}
\end{equation}
We adopted intrinsic optical axis ratios \(q_0=0.10\) and 0.15 independently of the gas scale heights. Table~\ref{tab:inclination-template-response} summarizes the paired source-level changes \(\Delta DM=DM(q=0.2)-DM(q=0.8)\). All DMs are in the host rest frame.

\begin{table}[h]
\centering
\caption{Inclination-induced DM changes in the electron-density templates.}
\label{tab:inclination-template-response}
\small
\begin{tabular*}{\textwidth}{@{\extracolsep{\fill}}lccccc@{}}
\hline\hline
& & \multicolumn{2}{c}{NE2001 + YT2020} & \multicolumn{2}{c}{NE2025 + YT2020} \\
\cline{3-4}\cline{5-6}
\(h_{\rm src}\) (pc) & \(q_0\) & Mean & 16th--84th percentiles & Mean & 16th--84th percentiles \\
\hline
330 & 0.10 & 180.8 & [65.9, 301.2] & 157.9 & [43.3, 265.3] \\
330 & 0.15 & 228.1 & [77.0, 384.1] & 193.7 & [47.7, 330.8] \\
100 & 0.10 & 187.3 & [85.8, 281.3] & 163.2 & [52.3, 252.9] \\
100 & 0.15 & 239.7 & [101.4, 367.8] & 202.7 & [58.3, 321.1] \\
\hline
\end{tabular*}
\par\smallskip
\begin{minipage}{\textwidth}
\footnotesize
\textit{Note.}---Values are host-rest-frame \(\Delta DM=DM(q=0.2)-DM(q=0.8)\), in \(\mathrm{pc\,cm^{-3}}\). Percentiles describe variation among source positions at fixed model parameters and viewing azimuth, not uncertainty in the population mean.
\end{minipage}
\end{table}
\end{document}